\pdfoutput=1
\documentclass[amsmath,prd,aps,twocolumn,superscriptaddress,floatfix,nofootinbib]{revtex4}

\usepackage{amsfonts}
\usepackage{bm}
\usepackage{color}
\usepackage{graphicx}
\usepackage{diagbox}
\definecolor{darkgreen}{cmyk}{0.85,0.2,1.00,0.2}

\newcommand{\be}{\begin{equation}}
\newcommand{\ee}{\end{equation}}
\newcommand{\ba}{\begin{eqnarray}}
\newcommand{\ea}{\end{eqnarray}}
\newcommand{\nn}{\nonumber}

\newcommand\lsim{\mathrel{\rlap{\lower4pt\hbox{\hskip1pt$\sim$}}
        \raise1pt\hbox{$<$}}}
\newcommand\gsim{\mathrel{\rlap{\lower4pt\hbox{\hskip1pt$\sim$}}
        \raise1pt\hbox{$>$}}}

\def\DM{\mbox{DM}}
\def\th{{\bm{\theta}}}
\def\x{{\bf x}}
\def\k{{\bf k}}
\def\l{{\bf l}}
\def\Si{\mbox{Si}}
\def\Ci{\mbox{Ci}}
\def\bN{{\bar N}}

\def\Cov{\mbox{Cov}}

\def\Plz{\,P_{\rm lin}\bigg(\frac{l}{\chi(z)},z\bigg)}
\def\plz{P_{\rm lin}(l/\chi(z),z)}

\def\Pgelz{\,P_{ge}\bigg(\frac{l}{\chi(z)},z\bigg)}

\begin{document}

\title{Characterizing fast radio bursts through statistical cross-correlations}

\author{Masoud Rafiei-Ravandi}
\affiliation{Perimeter Institute for Theoretical Physics, Waterloo, ON N2L 2Y5, Canada}
\author{Kendrick M.~Smith}
\affiliation{Perimeter Institute for Theoretical Physics, Waterloo, ON N2L 2Y5, Canada}
\author{Kiyoshi W. Masui}
\affiliation{MIT Kavli Institute for Astrophysics and Space Research, Cambridge, MA 02139, USA}
\affiliation{Department of Physics, Massachusetts Institute of Technology, Cambridge, MA 02139, USA}

\date{\today}


\begin{abstract}
Understanding the origin of fast radio bursts (FRB's) is a central unsolved
problem in astrophysics that is severely hampered by their poorly
determined distance scale.
Determining the redshift distribution of FRB's appears to require arcsecond angular resolution,
in order to associate FRB's with host galaxies.
In this paper, we forecast prospects for determining the redshift distribution
without host galaxy associations, by cross-correlating FRB's with a galaxy
catalog such as the SDSS photometric sample.
The forecasts are extremely promising: a survey such as CHIME/FRB that measures catalogs of $\sim 10^3$ FRB's
with few-arcminute angular resolution can place strong constraints on the FRB redshift distribution, by
measuring the cross-correlation as a function of galaxy redshift $z$ and FRB dispersion measure $D$.
In addition, propagation effects from free electron inhomogeneities modulate the observed
FRB number density, either by shifting FRB's between dispersion measure (DM) bins or
through DM-dependent selection effects.
We show that these propagation effects, coupled with the spatial clustering
between galaxies and free electrons,
can produce FRB-galaxy correlations which are comparable
to the intrinsic clustering signal.
Such effects can be disentangled based on their angular and $(z, D)$
dependence, providing an opportunity to study not only FRB's but the
clustering of free electrons.

\end{abstract}


\maketitle

\section{Introduction}

Fast radio bursts (FRB's) are an astrophysical transient
whose origin is not yet understood.  Since initial discovery in 2007~\cite{Lorimer:2007qn},
interest in FRB's has grown, and explaining the FRB phenomenon is now a central unsolved
problem in astrophysics (see~\cite{Katz:2016dti,Platts:2018hiy,Petroff:2019tty} for recent reviews).

An FRB is a short (usually 1--10 ms), bright ($\sim$1 Jy) radio pulse
which is highly dispersed: the arrival time at radiofrequency $\nu$ is
delayed, by an amount proportional to $\nu^{-2}$.
This dispersion relation arises naturally if the pulse propagates
through a cold plasma of free electrons.
In this case, the delay is proportional to the ``dispersion measure'' (DM),
which is defined as the electron column density along the line of sight:
\ba
\mbox{(Delay)} 
&=& (\DM) \left( \frac{e^2}{2\pi m_e c} \right) \nu^{-2} \\
&=& (4.15 \mbox{ ms}) \left( \frac{\DM}{1 \mbox{ pc cm}^{-3}} \right) \left( \frac{\nu}{1 \mbox{ GHz}} \right)^{-2}  \label{eq:dispersion}
\ea
where
\be
\DM \equiv \int n_e(x) \, dx \, .
\ee
FRB's are a population of dispersed pulses whose observed DM significantly
exceeds the maximum Galactic column density $\DM_{\rm gal}$ (inferred
from a model of the Galaxy~\cite{Cordes:2002wz,YKW}).
On most of the sky, $\DM_{\rm gal}$ is $\le 50$ pc cm$^{-3}$, and
FRB's are regularly observed with $\DM \gtrsim 1000$.
From the outset, the large DM suggested that FRB's were extragalactic,
although on its own the large DM could also be explained by a Galactic
event with a large local free electron density.
As more FRB's were observed, their sky distribution was found to be isotropic
(i.e.~not correlated with the Galactic plane), conclusively establishing an
extragalactic origin.

At the time of this writing, 92 FRB discoveries have been published
(according to FRBCAT~\cite{Petroff:2016tcr}, {\tt frbcat.org}).
Ten of these FRB's are ``repeaters'', meaning that multiple pulses have
been observed from the same source~\cite{Spitler:2016dmz,Scholz:2016rpt,Amiri:2019bjk,Andersen:2019yex}.
Nine of the repeaters were discovered by the CHIME/FRB instrument, and a
much larger sample of non-repeating FRB's from CHIME/FRB is expected soon.
(The authors are members of the CHIME/FRB collaboration, and forecasting the
scientific reach of CHIME/FRB was the main motivation for this paper.)

Determining the redshift distribution of FRB's is critical to understanding the
FRB phenomenon since a distance scale is required to determine the burst
energetics and volumetric rate.
In the next few paragraphs, we summarize the current observational status.

FRB's do not have spectral lines, so FRB redshifts cannot be directly determined.
When an FRB is observed, an upper bound on its redshift $z$ can be inferred from its
DM as follows.
We write the total DM of an FRB as the sum of contributions from our galaxy, 
the intergalactic medium (IGM), and the host galaxy:
\be
\DM = \DM_{\rm gal} + D_i(z) + D_h
\ee
where the IGM contribution is related to the FRB redshift as:
\be
D_i(z) = n_{e,0} \int_0^z dz' \, \frac{1+z'}{H(z')}  \label{eq:dm_igm}
\ee
where $n_{e,0}$ is the comoving electron number density and $H(z)$ is the Hubble expansion rate.
If we assume that $\DM_{\rm gal}$ is known precisely and subtracted, then the inequality $D_h \ge 0$
implies an upper bound on $z$.  A $\DM=1000$ FRB must satisfy $z \lesssim 0.95$, and a $\DM=3000$
FRB satisfies $z \lesssim 3.08$.
However, an alternative hypothesis is that FRB's are at much lower redshifts, and have large host DM's.

Three FRB's have been observed in long-baseline interferometers with sufficient
angular resolution to uniquely identify a host galaxy, and thereby determine a
redshift~\cite{Chatterjee:2017dqg,Marcote:2017wan,Tendulkar:2017vuq,Bannister:2019iju,Ravi:2019alc}.
The inferred redshifts are $z=0.19$, 0.32, and 0.66.
These observations suggest that most of the DM is IGM-related, but with only three
data points it cannot be concluded that this is true for the entire
population.

Host galaxy associations are a powerful way to determine FRB redshifts,
but require angular resolution around 1 arcsecond or better~\cite{Eftekhari:2017tbx}.
Unfortunately, most telescopes capable of finding large numbers of FRB's have angular
resolution much worse than this.
In particular, for most of the CHIME/FRB sources, the angular resolution is either
$\approx 1'$ or $\approx 10'$, depending on whether baseband data is available
for the event~\cite{kiyo_beamforming,Amiri:2018qsq,Andersen:2019yex}.

In this paper, we study the following question.
Given a catalog of FRB's whose resolution is insufficient for host galaxy
associations on a per-object basis, is it possible to associate FRB's and galaxies
on a {\em statistical} basis?
To make this question precise, we model the angular cross power spectrum $C_l^{fg}$
between the FRB and galaxy catalogs and forecast its signal-to-noise ratio (SNR).
The SNR turns out to be surprisingly large.
For example, given a catalog of 1000 FRB's with $1'$ resolution, and the
photometric galaxy catalog from SDSS-DR8~\cite{Aihara:2011sj}, we find an
SNR of 25--100, depending on the FRB redshift distribution.

As a consequence of this high SNR, the cross-correlation is still detectable
if the FRB and galaxy catalogs are binned in various ways.
By dividing the galaxy catalog into redshift bins, and separately 
cross-correlating each bin with the FRB catalog, the FRB redshift
distribution can be constrained.
By additionally dividing the FRB catalog into DM bins, the FRB redshift
distribution of each DM bin can be constrained, pinning down the
redshift-DM correspondence.

Other binning schemes are possible.
For example, the FRB catalog can be binned in observed flux,
so that the galaxy cross-correlation pins down the redshift-flux correspondence,
and therefore the intrinsic luminosity distribution of FRB's.
Or the galaxy catalog can be binned by star formation rate before cross-correlating
with FRB's, to determine whether FRB's are associated with star formation.
This technique can be applied easily to other tracer fields such as supernovae and
quasars.

This paper overlaps significantly with work in the galaxy clustering literature on
``clustering redshifts''~\cite{McQuinn:2013ib,Menard:2013aaa,Rahman:2014lfa,Kovetz:2016hgp,Passaglia:2017lnq}.
This term refers to the use of clustering statistics to determine the redshift distribution of 
a source population, by cross-correlating with a galaxy catalog.

However, in the case of FRB's, we find a significant new ingredient: large propagation effects,
which arise because galaxies are spatially correlated with free electrons, which in turn can
affect the observed density of FRB's and its DM dependence.
Propagation effects produce additional contributions to the FRB-galaxy angular correlation,
which need to be modeled and disentangled from the cosmological contribution.
In particular, if a galaxy catalog and an FRB catalog are correlated, this does not imply that
they overlap in redshift.
Propagation effects can also produce a correlation between low-redshift galaxies and high-redshift
FRB's (but not vice versa).
The propagation effects which we will explore have some similarity with magnification bias
in galaxy surveys (see e.g.~\cite{Hui:2007cu} and references therein).

We also clarify which properties of the FRB population are observable
via cross correlations.
It is well known that on large scales (``2-halo dominated'' scales), the only
observable is $(b_f dn_f/dz)$: the product of FRB redshift distribution $dn_f/dz$
and the large-scale clustering bias $b_f(z)$.
We find that there is an analogous observable $(\gamma_f dn_f/dz)$
which determines the FRB-galaxy correlation on smaller (``1-halo dominated'') scales.
The quantity $\gamma_f(z)$ measures the degree of similarity between the dark matter
halos which contain FRB's and galaxies, and is defined and discussed in~\S\ref{sec:power_spectrum_observables}.

This paper is complementary to previous works which have considered
different FRB-related clustering statistics.
In~\cite{Masui:2015ola}, the 3-d clustering statistics of the FRB field
were studied, using the DM as a radial coordinate. This is analogous to
the way photometric galaxy surveys are analyzed in cosmology.
Here we generalize to the cross correlation between the FRB field and a galaxy
survey.  The FRB-galaxy cross correlation has higher SNR than the FRB auto
correlation, since the number of galaxies is much larger than the number of
FRB's. Whereas~\cite{Masui:2015ola} was entirely perturbative, we perform both
perturbative calculations and non-linear simulations using a halo model. In
addition we consider two propagation effects: DM shifting and completeness (to
be defined below),
whereas \cite{Masui:2015ola} considered only the former.

Another idea that has been considered is to cross-correlate a 2-d map of FRB-derived dispersion measures
with galaxy catalogs, to probe the distribution of electrons in dark matter 
halos~\cite{McQuinn:2013tmc,Shirasaki:2017otr,Ravi:2018ose,Munoz:2018mll,Madhavacheril:2019buy}.
The cross-correlation of DM vs galaxy density is related to the DM moment of
the statistic $C_l^{fg}(z,D)$ considered here.  Therefore,
our statistic contains a superset of the information in the statistic
considered in these works.

In~\cite{Li:2019fsg}, a cross correlation was observed between
2MPZ galaxies at $z \sim 0.01$, and a sample of 23 FRB's from ASKAP
operating in ``fly-eye'' mode with $10'$--\,$60'$ angular resolution~\cite{Bannister:2017sie,2018Natur.562..386S}.
This measurement is seemingly at odds with the three FRB host galaxy redshifts
which imply a much more distant population.
In the very near future, FRB catalogs will be available with much higher number
density and better angular resolution, so it will be possible to measure the
cross correlation at higher SNR, and push the measurement to higher redshift.
The machinery in this paper will be essential for interpreting a high-SNR cross
correlation, and separating the clustering signal from propagation effects.

This paper is organized as follows.
In \S\ref{sec:preliminaries}, we define notation and our modeling assumptions.
In \S\ref{sec:clfg}, we define our primary observable, the FRB-galaxy cross power spectrum $C_l^{fg}$.
We explore and interpret clustering contributions to $C_l^{fg}$ in~\S\ref{sec:power_spectrum_observables},
and propagation effects in~\S\ref{sec:propagation_effects}.
We present signal-to-noise forecasts in \S\ref{sec:forecasts}, and in \S\ref{sec:simulations}
we describe a Monte Carlo simulation pipeline which we use to validate our forecasts.
We conclude in~\S\ref{sec:discussion}.

\section{Preliminaries}
\label{sec:preliminaries}

Throughout the paper, we use the flat-sky approximation, in which an angular
sky location is represented by a two-component vector $\th=(\theta_x,\theta_y)$,
and assume periodic boundary conditions with no angular mask for simplicity.
Angular wavenumbers are denoted $\l = (l_x,l_y)$, and 3-d comoving wavenumbers
are denoted $\k$.
We denote the observed sky area in steradians by $\Omega$.

Let $H(z)$ be the Hubble expansion rate at redshift $z$, and let $\chi(z)$ be
the comoving distance to redshift $z$:
\be
\chi(z) = \int_0^z \frac{dz'}{H(z')}
\ee
Let $P_{\rm lin}(k,z)$ denote the linear matter power spectrum at comoving
wavenumber $k$ and redshift $z$.

We use $f$ and $g$ to denote an FRB or galaxy catalog.
Depending on context, the FRB catalog may be binned in DM, or the galaxy catalog
may be binned in redshift.
For $X\in \{f,g\}$, let $n_X^{2d}$, $n_X^{3d}(z)$, and $dn_X^{2d}/dz$ denote the
2-d number density, 3-d number density, and 2-d number density per unit redshift.
These densities are related to each other by:
\be
n_X^{3d}(z) = \frac{H(z)}{\chi(z)^2} \frac{dn_X^{2d}}{dz}
\hspace{1cm}
n_X^{2d} = \int dz \, \frac{dn_X^{2d}}{dz}  \label{eq:X_densities}
\ee
We model FRB and galaxy clustering using the halo model.
For a review of the halo model, see~\cite{Cooray:2002dia}.
In this section, we give a high-level summary of our halo modeling formalism.
For details, see Appendix~\ref{app:halo_model}.

In the halo model, FRB and galaxy catalogs are simulated by a three-step process.
First, we simulate a random realization of the {\em linear} cosmological density field
$\delta_{\rm lin}(\th,z)$.  Since $\delta_{\rm lin}$ is a Gaussian field, its statistics
are completely determined by its power spectrum $P_{\rm lin}(k,z)$.

Second, we randomly place dark matter halos, which are modeled as biased Poisson
tracers of $\delta_{\rm lin}$.  More precisely, the probability of a halo in mass
range $(M,M+dM)$ and comoving volume $d^3\x$ near spatial location $\x$ is:
\be
n_h^{3d}(M,z) \Big( 1 + b_h(M,z) \delta_{\rm lin}(x) \Big) d^3\x \, dM
\ee
where $n_h^{3d}(M,z)$ is the {\em halo mass function}, or number density of
halos per unit comoving volume per unit halo mass, and $b_h(M,z)$ is the
{\em halo bias}.  We use the Sheth-Tormen mass function and bias
(Eqs.~(\ref{eq:sheth_tormen_halo_mass_function}),~(\ref{eq:sheth_tormen_halo_bias})).

Third, we randomly assign FRB's and galaxies to halos.
We always assume that the number counts $(N_f,N_g)$ of FRB's and galaxies are
independent from one halo to the next.
That is, $(N_f,N_g)$ is a bivariate random variable whose probability distribution
(the {\em halo occupation distribution} or HOD) depends only on halo mass $M$ and
redshift $z$.
Once the counts $(N_f,N_g)$ have been simulated, we assign spatial locations to
each FRB and galaxy independently, by sampling from the NFW spatial profile
(Eq.~(\ref{eq:nfw_real})).
We assume that galaxy positions are measured with negligible uncertainty,
but FRB positions have statistical errors $(\theta_x, \theta_y)$ which are
Gaussian with FWHM denoted $\theta_f$.
Unless stated otherwise, we take the FRB angular resolution to be
$\theta_f = 1$~arcminute.

\begin{figure*}
\centerline{
	\includegraphics[width=8.6cm]{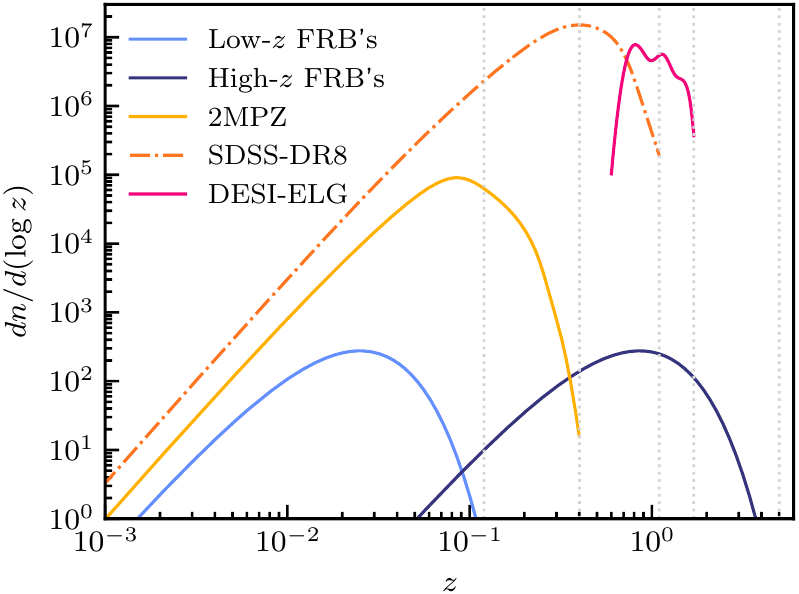}
	\hspace{0.2cm}
	\includegraphics[width=8.450434782608696cm]{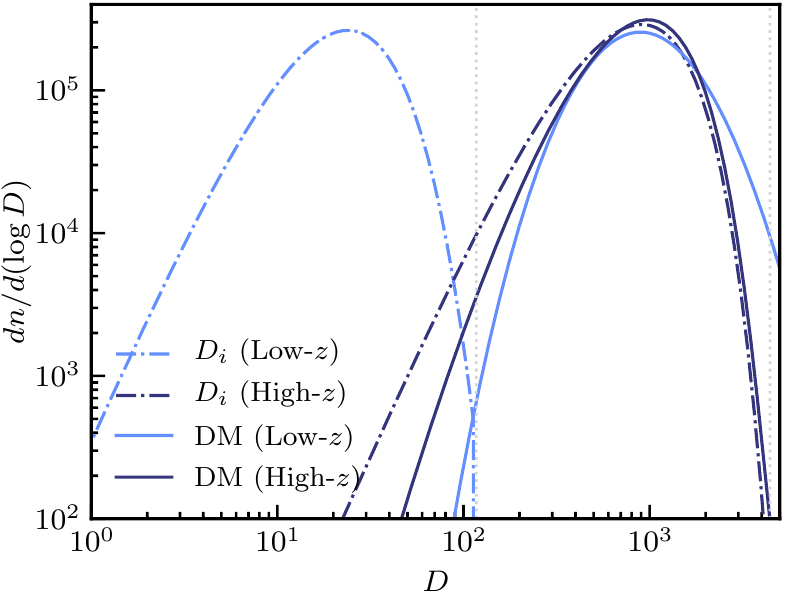}
}
\caption{{\em Left panel:} FRB redshift distributions in our high-$z$ and
low-$z$ fiducial FRB models (see \S\ref{sec:preliminaries}), with galaxy 
redshift distributions shown for comparison.
{\em Right panel:} FRB DM distributions in both fiducial models.
We show total extragalactic DM (IGM+host, denoted ``DM''), and the IGM contribution $D_i(z)$.
The total DM distribution is similar in the two fiducial models, but DM's are usually host-dominated
in the low-$z$ model, and IGM-dominated in the high-$z$ model.
Vertical dotted lines mark maximum redshift cutoffs.}
\label{fig:dndz}
\end{figure*}

Throughout the paper, we derive analytic results for an arbitrary HOD,
but show numerical results for two specific FRB models: the ``low-$z$'' and
``high-$z$'' fiducial FRB models.
Our two fiducial models are intended to bracket the range of possibilities
for the FRB redshift distribution currently allowed by observations.
The median FRB redshift in the low-$z$ and high-$z$ FRB models is
$z \sim 0.022$ and $z \sim 0.76$ respectively.
The {\em host} DM distributions in the two models have been chosen so that the
distribution of {\em total} DM's is similar (Figure~\ref{fig:dndz}).
In the high-$z$ FRB model, observed DM is a fairly good indicator of the FRB redshift,
whereas in the low-$z$ FRB model, there is not much correlation between DM and redshift.
The high-$z$ FRB model was motivated by the FRB host galaxy associations at redshifts
0.19, 0.32, 0.66 reported in~\cite{Chatterjee:2017dqg,Marcote:2017wan,Tendulkar:2017vuq,Bannister:2019iju,Ravi:2019alc},
and the low-$z$ FRB model was motivated by the ASKAP-2MPZ cross correlation at very
low redshift reported in~\cite{Li:2019fsg}.

In both FRB models, we define the FRB HOD so that FRB's have a small nonzero
probability to occur in halos above threshold mass $M_f = 10^9$ $h^{-1}$ $M_\odot$.
We have chosen $M_f$ to be small, roughly the minimum halo mass needed to host
a dwarf galaxy, since one FRB (the original repeater) is known to be in a dwarf.
If $M_f$ is increased (keeping the total number of observed FRB's $N_{\rm frb}$ fixed)
then the FRB-galaxy cross-correlations SNR also increases.
Therefore, our choice of small $M_f$ makes our forecasts a bit conservative.

We consider three galaxy surveys throughout the paper.
First, the SDSS-DR8 optical photometric survey over redshift
range $0 \le z \le 1.1$, with redshift distribution taken
from~\cite{Sheldon:2011fm}.
Second, the 2MPZ all-sky infrared photometric survey~\cite{Bilicki:2013sza},
Almost all ($\approx 98\%$) of the 2MPZ galaxies have photometric redshifts $< 0.2$.
Finally, the upcoming DESI-ELG spectroscopic survey, whose redshift distribution
is forecasted in~\cite{Aghamousa:2016zmz} and covers the range $0.6 \le z \le 1.7$.
For photometric surveys, we neglect photometric redshift uncertainties, since these
will be small compared to the FRB redshift uncertainty arising from scatter in the
FRB host DM.

\begin{figure}
\centerline{\includegraphics[width=8.6cm]{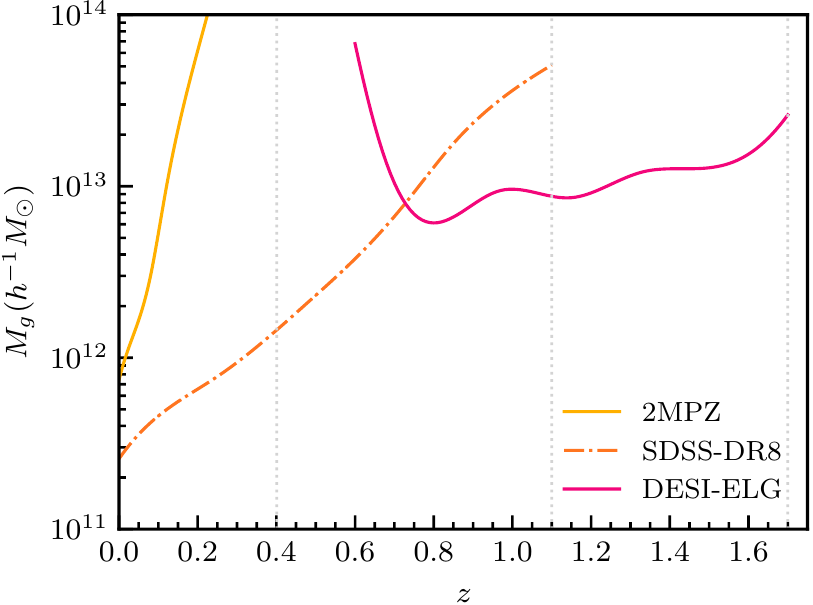}}
\caption{Threshold halo mass $M_g(z)$ for hosting a galaxy in the 2MPZ, SDSS-DR8 and
DESI-ELG galaxy surveys, determined by abundance matching to the redshift distribution
$dn_g/dz$ as described in \S\ref{sec:preliminaries} and Appendix~\ref{app:gal_hod}.
Vertical dotted lines mark maximum redshift cutoffs.}
\label{fig:mg}
\end{figure}

The galaxy HOD is constructed so that halos above threshold mass $M_g(z)$ contain
$(M/M_g(z))$ galaxies on average.
The redshift-dependent threshold halo mass $M_g(z)$ is chosen to match the redshift
distribution of the galaxy survey (``abundance matching'').
Numerical values of $M_g(z)$ are shown in Figure~\ref{fig:mg}.

For more details of the FRB and galaxy models, including precise specification of
the FRB redshift and host DM distributions in the two fiducial models, see
Appendices~\ref{app:gal_hod},~\ref{app:frb_hod}.

\section{The power spectrum $C_l^{fg}$}
\label{sec:clfg}

\subsection{Definition}

Our primary statistic for FRB-galaxy cross correlations
is the angular power spectrum $C_l^{fg}$, which measures
the level of correlation as a function of angular wavenumber $l$.

We review the definition of the angular power spectrum.
The input data is a catalog of FRB sky locations $\th^f_1, \cdots, \th^f_{N_f}$,
and a catalog of galaxy sky locations $\th^g_1, \cdots, \th^g_{N_g}$.
We then define the 2-d FRB field $\delta_f(\th)$ as a sum of delta functions:
\be
\delta_f(\th) = \frac{1}{n_f^{2d}} \sum_{i=1}^{N_f} \delta^2(\th-\th_i^{(f)})
\ee
and similarly for the galaxy field $\delta_g(\th)$.

In Fourier space, the FRB field $\delta_f(\l)$ is a sum of complex exponentials:
\be
\delta_f(\l) = \frac{1}{n_f^{2d}} \sum_{i=1}^{N_f} \exp\Big( -i\l\cdot\th_i^{(f)} \Big)
\ee
and likewise for $\delta_g$.  The two-point correlation function of the fields 
$\delta_f, \delta_g$ is simplest in harmonic space, where it takes the form:
\be
\langle \delta_f(\l)^* \, \delta_g(\l') \rangle = C_l^{fg} (2\pi)^2 \delta^2(\l-\l')
\ee
where the delta function on the RHS is a consequence of translation invariance.
This equation defines the power spectrum $C_l^{fg}$.

The power spectrum $C_l^{fg}$ is one representation for the two-point correlation
function between $\delta_f, \delta_g$.
Other representations, such as the two-point correlation function as a function of
angular separation, contain the same information as $C_l^{fg}$.
The power spectrum has the advantage that when it is estimated from data, statistical
correlations between different $l$-values are small (in contrast with the correlation
function, where correlations between different angular separations can be large).
For this reason, we choose to use the angular power spectrum throughout the paper.

If the galaxy catalog has been divided into redshift bins,
then for each redshift bin $j$ we can define a galaxy field $\delta_{g_j}(\th)$,
and a power spectrum $C_l^{fg_j}$ by cross-correlating with the (unbinned)
FRB catalog.

Similarly, we can bin the FRB's by dispersion measure.
Throughout the paper, we assume that the galactic contribution $\DM_{\rm gal}$ 
can be accurately modeled, and subtracted from the observed DM prior to binning.
For each FRB DM bin $i$ and galaxy redshift bin $j$, we can compute an angular
power spectrum $C_l^{f_ig_j}$.
In the limit of narrow redshift and DM bins, the angular power spectrum becomes
a function $C_l^{fg}(z,D)$ of three variables: angular wavenumber $l$, galaxy redshift $z$, and
FRB dispersion measure $D$.

\subsection{Two-halo and one-halo power spectra}
\label{ssec:clfg_1h_2h}

In the halo model, the power spectrum $C_l^{fg}$ can be calculated exactly.
Here we summarize the main features of the calculation; details are in Appendix~\ref{app:halo_model}.

\begin{figure*}
\centerline{
  \includegraphics[width=8.6cm]{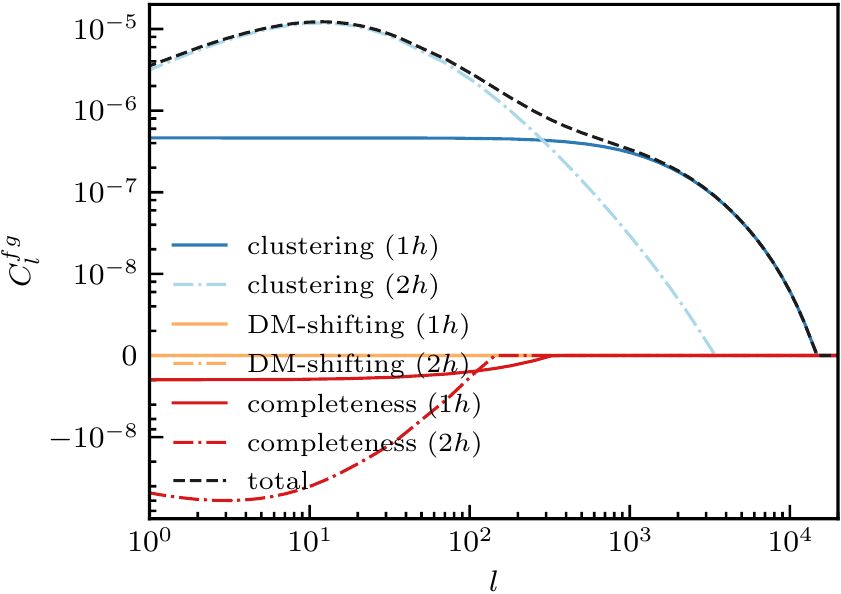}
  \hspace{0.42cm}
  \includegraphics[width=8.6cm]{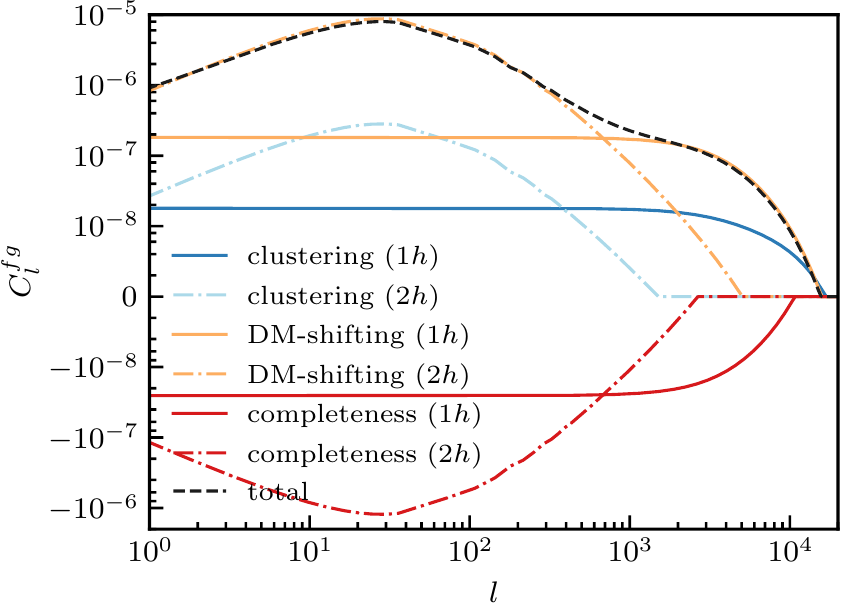}
}
\caption{Angular cross power spectrum $C_l^{fg}$ as a function of $l$ for the high-$z$ fiducial FRB model (see~\S\ref{sec:preliminaries}) 
  and SDSS-DR8 galaxies.  The total observed power spectrum is the sum of clustering and propagation contributions, and each contribution
  may be split into 1-halo and 2-halo terms, which we show separately here.
  Disentangling these terms is a challenge, and one of the main themes of this paper.
  The clustering terms are described in~\S\ref{ssec:clfg_1h_2h}, and
  the ``DM-shifting'' and ``completeness'' terms are propagation effects which will be described in~\S\ref{sec:propagation_effects}.
  {\em Left panel:} unbinned FRB and galaxy fields.
  {\em Right panel:} FRB dispersion measure bin $1400 < D < 1500$ and galaxy redshift bin $0.63 < z < 0.74$.}
\label{fig:clfg_l}
\end{figure*}

The power spectrum is the sum of {\em 2-halo} and {\em 1-halo} terms:
\be
C_l^{fg} = C_l^{fg(2h)} + C_l^{fg(1h)}
\ee
which correspond to correlations between FRB's and galaxies in different halos,
or in the same halo.
Some example 2-halo and 1-halo power spectra are shown in Figure~\ref{fig:clfg_l}.

The 2-halo term $C_l^{fg(2h)}$ is sourced by large-scale
cosmological correlations, and is responsible for the large bump
at low $l$.
For an arbitrary redshift $z$, the bump is at $l \sim k_{\rm eq} \chi(z)$, where
$k_{\rm eq} \sim 0.02$ $h$ Mpc$^{-1}$ is the scale of matter-radiation equality.
The 2-halo term arises because FRB's and galaxies trace the same underlying large-scale
cosmological density fluctuations.
On large scales (low $l$), where halo profiles and beam resolution are negligible,
$C_l^{fg(2h)}$ takes the form:
\ba
C_l^{fg(2h)} & \rightarrow & \frac{1}{n_f^{2d} n_g^{2d}} \int dz \, \frac{H(z)}{\chi(z)^2} \left( b_f(z) \frac{dn_f^{2d}}{dz} \right) \nn \\
 && \hspace{0.5cm} \times \left( b_g(z) \frac{dn_g^{2d}}{dz} \right) \Plz  \label{eq:clfg_2h_lowl}
\ea
(For a more precise expression for $C_l^{fg(2h)}$ which applies at high $l$, see Eq.~(\ref{eq:clfg_master}) in Appendix~\ref{app:halo_model}.)

\begin{figure}
\centerline{\includegraphics[width=8.6cm]{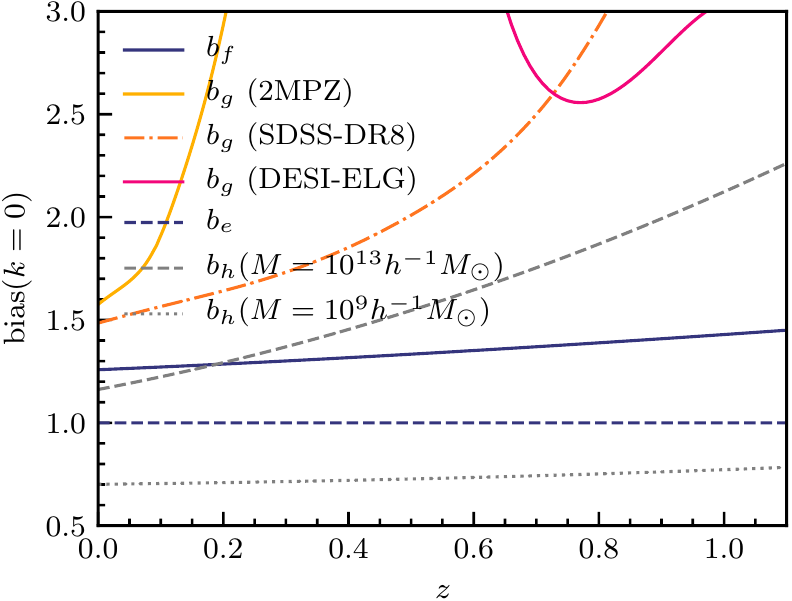}}
\caption{Large-scale bias parameters.  The FRB bias $b_f(z)$ assumes minimum halo mass $M_f = 10^9$ $h^{-1}$ $M_\odot$.
 The galaxy bias $b_g(z)$ for the 2MPZ, SDSS-DR8 and DESI-ELG surveys assumes the minimum halo mass $M_g(z)$ from Figure~\ref{fig:mg}.
 We take electron bias $b_e=1$ throughout.
 We also show the halo bias $b_h(z)$ for two choices of halo mass.}
\label{fig:biases}
\end{figure}

Here, $b_f(z), b_g(z)$ are bias parameters which measure the coupling
of FRB's and galaxies to the cosmological density field on large scales.
The FRB bias $b_f$ is defined by the statement that the FRB and matter
overdensities are related by $\delta_f \approx b_f \delta_m$ on large
scales, and likewise for $b_g$.
An explicit formula for $b_f,b_g$ is given in Eq.~(\ref{eq:bias_X}).
and numerical values are shown in Figure~\ref{fig:biases}.
The 2-halo term mainly depends on the redshift overlap between the FRB and galaxy
catalogs, via the factors $(b_f dn_f^{2d}/dz) (b_g dn_g^{2d}/dz)$ in Eq.~(\ref{eq:clfg_2h_lowl}).

The 1-halo term $C_l^{fg(1h)}$ arises because FRB's and galaxies
occupy the same dark matter halos.
On large scales (low $l$), where halo profiles and beam resolution are negligible,
the 1-halo term takes the form:
\be
C_l^{fg(1h)} \rightarrow \frac{1}{n_f^{2d} n_g^{2d}} \int dz \, dM \, \frac{\chi(z)^2}{H(z)} n_h^{3d}(M,z) \, \big\langle N_f N_g \big\rangle_{M,z}  \label{eq:clfg_1h_lowl}
\ee
where $\langle \cdot \rangle_{M,z}$ denotes the average over the HOD
in a halo of mass $M$ at redshift $z$.
(For a more precise expression for $C_l^{fg(1h)}$ which applies at high $l$, see Eq.~(\ref{eq:clfg_master}) in Appendix~\ref{app:halo_model}.)

The 1-halo term is harder to interpret than the 2-halo term, since
it depends on the details of the HOD.
As an artificial example, suppose that the FRB and galaxy catalogs
do overlap in redshift, but the FRB and galaxy HOD's do not overlap
in halo mass.
Then the 1-halo term will be zero.  This example is artificial,
since halos of sufficiently large mass will contain galaxies of all types,
and presumably FRB's as well.  However, it illustrates that interpreting
the 1-halo term is not straightforward.  We will return to this issue
shortly.

The 1-halo term $C_l^{fg(1h)}$ arises whenever FRB's and survey galaxies occupy the same halos.
If FRB's actually inhabit the survey galaxies themselves, there will be an additional ``Poisson'' term $C_l^{fg(p)}$
which dominates on the smallest scales (high $l$).
We have neglected the Poisson term in our forecasts, since we are assuming that the FRB survey has insufficient
resolution to associate FRB's and galaxies on a per-object basis, but this does make our forecasts slightly conservative.
For more discussion of the Poisson term, see Eq.~(\ref{eq:clfg_poisson}) in Appendix~\ref{app:halo_model}.

\section{The observables $b(dn/dz)$ and $\gamma(dn/dz)$}
\label{sec:power_spectrum_observables}

In the limit of narrow galaxy redshift and FRB DM bins,
the angular power spectrum $C_l^{fg}(z,D)$ is a function of three variables:
angular wavenumber $l$, FRB dispersion measure $D$, and galaxy redshift $z$.
One may wonder whether the information in $C_l^{fg}$ can be ``compressed''
into a function of fewer variables.

In this section, we will take a step in this direction, by showing how the
$l$-dependence can be absorbed into two observables, corresponding to the power
spectrum amplitude in the 2-halo and 1-halo regimes.
These observables, denoted $b(dn/dz)$ and $\gamma(dn/dz)$ for reasons
to be explained shortly, will be functions of $z$ and $D$.

The basic idea is simple.
For a narrow galaxy redshift bin $(z, z + \Delta z)$, the 2-halo and 1-halo power spectra
in Eqs.~(\ref{eq:clfg_2h_lowl}),~(\ref{eq:clfg_1h_lowl}) have the following limiting forms
at low~$l$:
\ba
C_l^{fg(2h)} & \rightarrow & \big( \mbox{Constant} \big) \Plz \nn \\
C_l^{fg(1h)} & \rightarrow & \big( \mbox{Constant} \big) \label{eq:clfg_coeffs}
\ea
At higher values of $l$, the power spectra acquire additional $l$-dependence which gives
information about halo profiles, but we will assume that this profile information is of
secondary interest.
Thus, the information in the $l$-dependence of the power spectrum can be compressed
into two numbers: the coefficients in Eq.~(\ref{eq:clfg_coeffs}).
Given a measurement of the total power spectrum $C_l^{fg}$, we can fit for
both coefficients jointly, without much covariance between them.

Starting with the 2-halo power spectrum, we take Eq.~(\ref{eq:clfg_2h_lowl}) in
the limit of a narrow redshift bin $(z,z+\Delta z)$, obtaining:
\be
C_l^{fg(2h)} \rightarrow \frac{1}{n_f^{2d}} \frac{H(z)}{\chi(z)^2} \left( b_f(z) \frac{dn_f^{2d}}{dz} \right) b_g(z) \Plz  \label{eq:clfg_2h_b}
\ee
All factors on the RHS are known in advance except $b_f(z) dn_f^{2d}/dz$, including the
factor $\plz$ which determines the $l$-dependence.
In particular, the galaxy bias $b_g(z)$ can be measured in several ways, for example by cross-correlating
the redshift-binned galaxy catalog with CMB lensing.
Therefore, we can interpret the 2-halo power spectrum amplitude as a measurement of the
quantity $b_f (dn_f^{2d}/dz)$.

The observable quantity $b_f (dn_f^{2d}/dz)$ is not as intuitive as the FRB redshift distribution $(dn_f^{2d}/dz)$,
but in practice the two are not very different.  For example, in our fiducial model with threshold halo
mass $M_f=10^9$ $h^{-1}$ $M_\odot$, the FRB bias satisfies $1.2 \le b_f \le 1.5$ for $z \le 1$
(see Figure~\ref{fig:biases}).

This interpretation of the 2-halo amplitude as a measurement of $b (dn/dz)$ is fairly standard and has been explored
elsewhere~\cite{McQuinn:2013ib,Menard:2013aaa,Rahman:2014lfa,Kovetz:2016hgp,Passaglia:2017lnq}.
The 1-halo amplitude is less straightforward to interpret, and does not seem to have a standard interpretation in the literature.
In the rest of this section, we will define an analogous observable $\gamma (dn/dz)$ for the 1-halo amplitude.
The definition is not specific to FRB's, and may be interesting in the context of other tracer populations.

We define the following 3-d densities:
\ba
n_{gg}^{3d}(z) &=& \int dM \, n_h^{3d}(M,z) \, \big\langle N_g^2 \rangle_{M,z} \label{eq:ngg_def} \\
n_{fg}^{3d}(z) &=& \int dM \, n_h^{3d}(M,z) \, \big\langle N_f N_g \rangle_{M,z} \label{eq:nfg_def}
\ea
where $\langle \cdot \rangle_{M,z}$ is the expectation value over the HOD for a halo of mass $M$ at redshift $z$.
These can be interpreted as comoving densities of pair counts $(g,g')$ or $(f,g)$ in the same halo.
Next we define:
\be
\gamma_f(z) = \frac{n_g^{3d}(z)}{n_f^{3d}(z)} \frac{n_{fg}^{3d}(z)}{n_{gg}^{3d}(z)}  \label{eq:gamma_def}
\ee
We will see shortly that the 1-halo amplitude can be interpreted as a measurement
of $\gamma_f (dn_f^{2d}/dz)$.

We would like to give an intuitive interpretation of $\gamma_f(z)$.
First, note that $\gamma_f$ is invariant under rescaling the overall abundance of FRB's and galaxies.
For example, if we wait until the FRB experiment has detected twice as many FRB's,
then densities rescale as $n_{fg}^{3d} \rightarrow 2 n_{fg}^{3d}$ and $n_f^{3d} \rightarrow 2 n_f^{3d}$,
leaving $\gamma_f$ unchanged.

Second, note that if the galaxy and FRB HOD's were identical (aside from overall abundance), then $\gamma_f(z) = 1$.
If the FRB HOD were then modified so that FRB's are in more massive halos (relative to the galaxies),
then $n_{fg}^{3d}$ would increase, and $\gamma_f(z)$ will be $> 1$.
Conversely, if the typical FRB inhabits a halo which is less massive than a typical galaxy,
then $\gamma_f(z)$ will be $< 1$.

\begin{figure}
\centerline{\includegraphics[width=8.6cm]{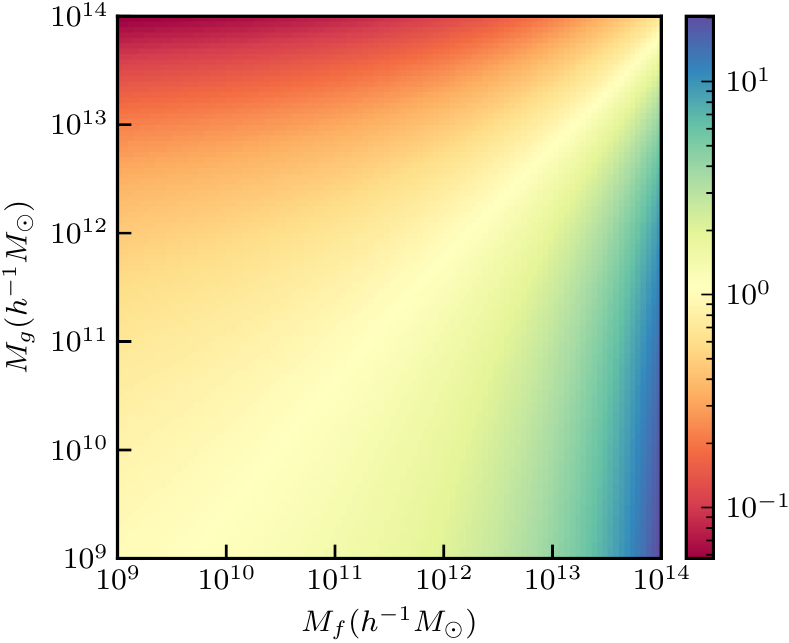}}
\vspace{0.275cm}
\centerline{\includegraphics[width=8.6cm]{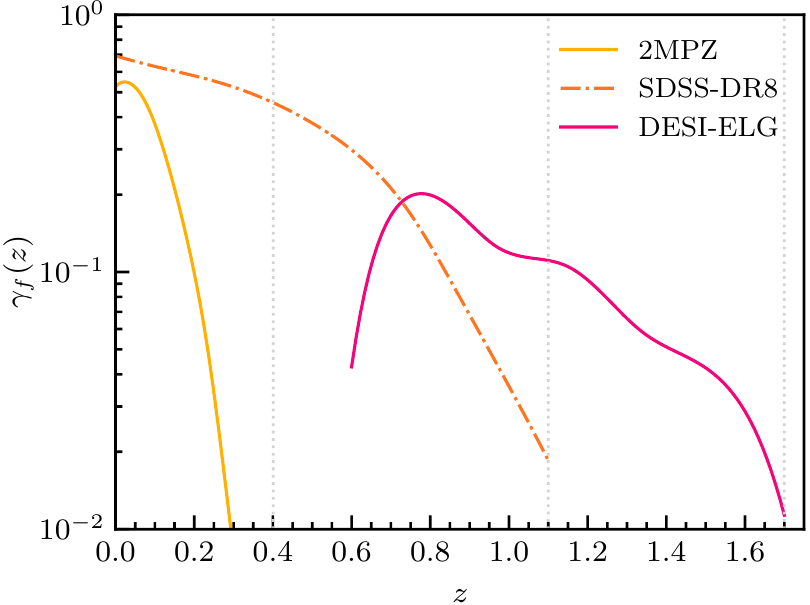}}
\caption{{\em Top panel:} Quantity $\gamma_f(z)$ defined in Eq.~(\ref{eq:gamma_def}), as a function of
  threshold FRB halo mass $M_f$ and threshold galaxy mass $M_g$, for Poisson HOD's at redshift $z=0.5$.
  If $M_f$ and $M_g$ are comparable, then $\gamma_f$ is of order 1.  {\em Bottom panel:} Quantity $\gamma_f(z)$
  as a function of redshift, assuming FRB threshold halo mass $M_f = 10^9$ $h^{-1}$ $M_\odot$, and galaxy
  threshold halo mass $M_g(z)$ from Figure~\ref{fig:mg}.
  At high redshifts, $\gamma_f$ can be $\ll 1$ in our models, since galaxies are rare and our abundance-matching
  prescription gives a large value of $M_g$.
  Vertical dotted lines mark maximum redshift cutoffs.}
\label{fig:gamma}
\end{figure}

In Figure~\ref{fig:gamma}, we show $\gamma_f(z)$ for our fiducial HOD (Eqs.~(\ref{eq:gal_hod}),~(\ref{eq:frb_hod}))
as a function of $(M_f,M_g)$, the threshold halo masses for FRB's and galaxies.
Consistent with the previous paragraph, if $M_f$ and $M_g$ are of the same order of magnitude, then $\gamma_f$
is of order unity.
In the regimes $M_f \ll M_g$ and $M_f \gg M_g$, the quantity $\gamma_f$ will be $\lesssim 1$ and $\gtrsim 1$
respectively.

Now we show how the 1-halo amplitude can be interpreted as a measurement of $\gamma_f(z) (dn_f^{2d}/dz)$.
We take Eq.~(\ref{eq:clfg_1h_lowl}) and specialize to a narrow redshift bin $(z, z + \Delta z)$, obtaining:
\be
C_l^{fg(1h)} \rightarrow \frac{1}{n^{2d}_f} \frac{n_{fg}^{3d}(z)}{n_g^{3d}(z)}  \label{eq:clfg_1h_lowl_narrowz}
\ee
Similarly, the 1-halo amplitude of the galaxy auto power spectrum is:
\be
C_l^{gg(1h)} \rightarrow \frac{1}{n_g^{2d}} \frac{n_{gg}^{3d}(z)}{n_g^{3d}(z)}
\ee
by specializing Eq.~(\ref{eq:clfg_master}) for $C_l^{gg(1h)}$ in Appendix~\ref{app:halo_model}
to low $l$ and a narrow redshift bin.
Now we write $C_l^{fg(1h)}$ in the following form:
\ba
C_l^{fg(1h)}
  & \rightarrow & \frac{n_g^{2d}}{n_f^{2d}} \gamma_f(z) \frac{n_f^{3d}(z)}{n_g^{3d}(z)} C_l^{gg(1h)} \nn \\
  &=& \frac{\Delta z}{n_f^{2d}} \left( \gamma_f(z) \frac{dn_f^{2d}}{dz} \right) C_l^{gg(1h)}  \label{eq:clfg_1h_gamma}
\ea
where the second line follows from the first by using Eq.~(\ref{eq:X_densities}).
All factors on the RHS are known in advance except $\gamma_f(z) dn_f^{2d}/dz$, including the
factor $C_l^{gg(1h)}$ which can be measured from the galaxy auto power spectrum.
Therefore, the 1-halo amplitude can be interpreted as a measurement
of the quantity $\gamma_f(z) dn_f^{2d}/dz$.

Summarizing, we have defined power spectrum observables $b_f (dn_f^{2d}/dz)$ and $\gamma_f (dn_f^{2d}/dz)$.
By measuring the power spectrum $C_l^{fg}$ as a function of $(l,z)$, both observables may be constrained
as functions of $z$.
This extracts all information in $C_l^{fg}$, except for suppression at high $l$ which contains information
about halo profiles.
The FRB catalog may be further binned in DM, to measure the observables $b_f (dn_f^{2d}/dz)$ and
$\gamma_f (dn_f^{2d}/dz)$ as functions of $(D,z)$.
In the top rows of Figures~\ref{fig:obs_high_z},~\ref{fig:obs_low_z}, we show the observables as functions
of $(D,z)$ in our fiducial model.

\section{Propagation Effects}
\label{sec:propagation_effects}

So far, we have considered contributions to $C_l^{fg}$ which arise because 3-d positions of
FRB's and galaxies are spatially correlated.
However, propagation effects also contribute to $C_l^{fg}$.
Galaxies at redshift $z_g$ will spatially correlate with free electrons, which
can modulate the observed abundance of FRB's at redshifts $z_f > z_g$, via dispersion,
scattering, or lensing.
This generates new contributions to $C_l^{fg}$, which we will study systematically in this section.

Throughout this section, $f$ denotes an FRB catalog, which may be constructed by selecting on FRB properties.
For example, $f$ could be a subcatalog of a larger catalog, obtained by selecting a DM bin or a fluence bin.

\subsection{Generalities}

Let $\delta_e(\th,z)$ be the 3-d electron overdensity along the past lightcone.
We will expand propagation effects to first order in $\delta_e$.

Let $\delta_f(\th)$ be the 2-d FRB overdensity produced by propagation effects,
given a realization of $\delta_e$.
We write $\delta_f$ as a line-of-sight integral:
\be
\delta_f(\th) = \int dz \, W_f(z) \delta_e(\th,z)  \label{eq:Wf_def}
\ee
where this equation defines the ``window function'' $W_f(z)$.
We will show how to calculate $W_f(z)$ shortly.

Given the window function $W_f(z)$, the contribution to $C_l^{fg}$ due to
propagation effects may be calculated from Eq.~(\ref{eq:Wf_def}).
In the Limber approximation, the result is:
\be
C_l^{fg} = \frac{1}{n_g^{2d}} \int dz \, W_f(z) n_g^{3d}(z) \Pgelz  \label{eq:clfg_p}
\ee
where $P_{ge}(k,z)$ is the 3-d galaxy-electron power spectrum at comoving wavenumber $k$.
We model $P_{ge}$ using the halo model (Eq.~(\ref{eq:pge})) in Appendix~\ref{app:halo_model}).
For a narrow galaxy redshift bin $(z,z+\Delta z)$, Eq.~(\ref{eq:clfg_p}) becomes:
\be
C_l^{fg} \rightarrow \frac{H(z)}{\chi(z)^2} W_f(z) \Pgelz  \label{eq:clfg_p_zbin}
\ee

\subsection{Dispersion-induced clustering}

In this section we will compute the window function $W_f(z)$ defined by Eq.~(\ref{eq:Wf_def}).
There will be contributions to $W_f(z)$
from several propagation effects: dispersion, scattering, and lensing.
In this paper, we will describe the dispersion case in detail, deferring
the other cases to future work.

For an FRB at sky location $\th$ and redshift $z_f$, we write the DM as $D = D_i(z_f) + \Delta(\th,z_f)$,
where $\Delta(\th,z_f)$ is the DM perturbation due to electron anisotropy along the line of sight at redshifts $0 < z < z_f$.
Then $\Delta$ is given explicitly by:
\be
\Delta(\th,z_f) = n_{e,0} \int_0^{z_f} dz \, \frac{1+z}{H(z)} \delta_e(\th,z)  \label{eq:Delta_DM}
\ee
As usual, let $dn_f^{2d}/dz$ denote the angular number density per unit redshift, so that:
\be
n_f^{2d} = \int dz \, \frac{dn_f^{2d}}{dz}  \label{eq:nf2d_z}
\ee
We introduce the notation $(\partial/\partial\Delta) (dn_f^{2d}/dz)$
to denote the derivative of $dn_f^{2d}/dz$ with respect to a foreground
DM perturbation $\Delta(z)$ along the line of sight.
Then, by differentiating Eq.~(\ref{eq:nf2d_z}), we can formally write the
propagation-induced FRB anisotropy as:
\be
\delta_f(\th) = \frac{1}{n_f^{2d}} \int dz_f \, \Delta(\th,z_f) \left( \frac{\partial}{\partial\Delta} \frac{dn_f^{2d}}{dz_f} \right) \nn \\
\ee
Plugging in Eq.~(\ref{eq:Delta_DM}) for $\Delta(\th,z_f)$ and reversing the order of integration, we get:
\be
\delta_f(\th)
 = \frac{n_{e,0}}{n_f^{2d}} \int dz \frac{1+z}{H(z)} \delta_e(\th,z)
     \int_z^\infty dz_f \, \left( \frac{\partial}{\partial\Delta} \frac{dn_f^{2d}}{dz_f} \right)
\ee
Comparing with the definition of $W_f$ in Eq.~(\ref{eq:Wf_def}) we read off the window function:
\be
W_f(z) = \frac{n_{e,0}}{n_f^{2d}} \, \frac{1+z}{H(z)} \int_z^\infty dz' \, \left( \frac{\partial}{\partial\Delta} \frac{dn_f^{2d}}{dz'} \right)  \label{eq:Wf_dispersion}
\ee
This identity relates the window function $W_f$ to the derivative $(\partial/\partial\Delta) (dn_f^{2d}/dz)$,
but it remains to compute the latter quantity.
This will depend on the details of how the FRB catalog $f$ is selected.

Generally speaking, the derivative $(\partial/\partial\Delta) (dn_f^{2d}/dz)$ contains two
terms.  First, there is a term which arises because a DM perturbation changes the probability
that an FRB is detected.  Increasing DM preserves pulse fluence, but decreases
signal-to-noise.\footnote{This is true for FRB searches based on incoherent
  dedispersion, such as the CHIME/FRB real-time search, due to pulse broadening within
  each frequency channel.  If the FRB search were based on coherent dedispersion,
  then dispersion would not change the SNR.  However, a coherent search is computationally
  infeasible for large blind searches.}
If the FRB catalog is constructed by selecting all objects above a fixed SNR threshold,
then this effect gives a negative contribution to $(\partial/\partial\Delta) (dn_f^{2d}/dz)$.
We will refer to this contribution as the {\em completeness} term.

Second, in the case where the FRB catalog is DM-binned, there is an additional term in
$(\partial/\partial\Delta) (dn_f^{2d}/dz)$ which arises because a DM perturbation can shift
observed DM's across a bin boundary.  We will refer to this contribution as the
{\em DM-shifting} term.

We give an explicit formula for the DM-shifting term as follows.
Suppose that the FRB catalog is constructed by selecting FRB's in DM bin $(D_{\rm min}, D_{\rm max})$.
Let $(d^2n_f^{2d} / dz \, dD)$ be the angular number density of FRB's per (redshift, DM), so that:
\be
\frac{dn_f^{2d}}{dz} = \int_{D_{\rm min}}^{D_{\rm max}} dD \, \frac{d^2n_f^{2d}}{dz \, dD}
\ee
Then the DM-shifting term is:
\be
\left( \frac{\partial}{\partial\Delta} \frac{dn_f^{2d}}{dz} \right)_{\rm DM\textnormal{-}shifting}
= \left( \frac{d^2n_f^{2d}}{dz \, dD} \right)_{D_{\rm min}} \hspace{-0.3cm} - \hspace{0.1cm} \left( \frac{d^2n_f^{2d}}{dz \, dD} \right)_{D_{\rm max}}  \label{eq:dm_shifting}
\ee
Next we give an explicit formula for the completeness term.
This term is more complicated and depends on both selection and the underlying FRB population.
As a toy model for exploring the order of magnitude of this term, we will make the following assumptions:
\begin{enumerate}
  \item The FRB catalog is constructed by selecting all objects above threshold signal-to-noise SNR$_*$.
  \item All FRB's have the same intrinsic pulse width $t_i$.
  \item In each redshift and DM bin, the FRB luminosity function is Euclidean: the number of FRB's above
    fluence $F_*$ is proportional to $(F_*^{-3/2})$.\footnote{The luminosity function is expected to
    be Euclidean at low $z$ if the FRB catalog is unbinned in redshift.  However, within a ($z$, DM)
    bin, there is no particular reason why the FRB luminosity function should be Euclidean, so this
    assumption of our toy model is fairly arbitrary.}
  \item SNR is related to fluence $F$ by
    \be
    \mbox{SNR} \propto \frac{F}{(t_i^2 + t_s^2 + t_d^2)^{1/4}}  \label{eq:snr_f}
    \ee
    where $t_s$ is the instrumental time sample length, and $t_d$ is the dispersion delay within a channel,
    given by
    \be
    t_d = 2 \mu (\mbox{DM}) \nu^{-3} (\Delta\nu)  \label{eq:td_def}
    \ee
    where $\nu$ is the observing frequency, $(\Delta\nu)$ is the channel bandwidth, and $\mu = 4.15$ ms GHz$^2$ is the
    coefficient in the FRB dispersion relation (delay) = $\mu (\mbox{DM}) / \nu^2$ in Eq.~(\ref{eq:dispersion}).
\end{enumerate}
Under these assumptions, we can calculate the derivative of $\log d^2n_f/(dz\,dD)$ 
with respect to a foreground DM perturbation $\Delta$, as follows:
\ba
\frac{\partial}{\partial\Delta} \left( \log \frac{d^2n_f}{dz\, dD} \right)
&=& -\frac{3}{2} \frac{\partial\log F_*}{\partial\Delta} \nn \\
&=& -\frac{3}{2} \frac{\partial\log (t_i^2+t_s^2+t_d^2)^{1/4}}{\partial\Delta} \nn \\
&=& -\frac{3t_d}{4(t_i^2+t_s^2+t_d^2)} \frac{\partial t_d}{\partial\Delta} \nn \\
&=& -\frac{3\mu (\Delta\nu) t_d}{2\nu^3 (t_i^2+t_s^2+t_d^2)}  \label{eq:completeness_2d}
\ea
Here, the first line follows from toy model assumption 3, the second line follows from Eq.~(\ref{eq:snr_f}),
the and the last line follows from differentiating Eq.~(\ref{eq:td_def}) with respect to DM.

To get the completeness term in the derivative $(\partial/\partial\Delta) (dn_f^{2d}/dz)$,
we integrate Eq.~(\ref{eq:completeness_2d}) over $D$:
\ba
&& \left( \frac{\partial}{\partial\Delta} \frac{dn_f^{2d}}{dz} \right)_{\rm completeness} \nn \\
&& \hspace{0.5cm} = \int dD \, \left( \frac{\partial}{\partial\Delta} \frac{d^2n_f^{2d}}{dz \, dD} \right) \nn \\
&& \hspace{0.5cm} = \int dD \, \left( \frac{d^2n_f^{2d}}{dz\,dD} \right) \left( -\frac{3\mu(\Delta\nu)t_d}{2\nu^3(t_i^2+t_s^2+t_d^2)} \right)  \label{eq:completeness}
\ea
In our toy model, the completeness term always gives a negative contribution to $C_l^{fg}$,
since increasing the DM of an FRB (at fixed fluence) decreases SNR.
This is true under the assumptions of our toy model, but is not guaranteed to be true in general.
For example in the CHIME/FRB real-time search, the RFI removal pipeline includes a filtering operation
which detrends intensity data along its radiofrequency axis, removing signal from low-DM events.
In principle this gives a positive contribution to $C_l^{fg}$, although end-to-end simulations of
the CHIME/FRB triggering pipeline would be needed to determine whether the overall sign is positive 
or negative.

Summarizing, in this section we have calculated two contributions to $C_l^{fg}$
from propagation effects: a ``DM-shifting'' term and a ``completeness'' term.
In both cases, the contribution to $C_l^{fg}$ is calculated as follows.
We compute the intermediate quantity $(\partial/\partial\Delta) (dn_f^{2d}/dz)$
using Eq.~(\ref{eq:dm_shifting}) or Eq.~(\ref{eq:completeness_2d}),
then the window function $W_f(z)$ using Eq.~(\ref{eq:Wf_dispersion}),
and finally $C_l^{fg}$ using Eq.~(\ref{eq:clfg_p}).

Finally, other studies have proposed to isolate these propagation effects to
measure $P_{ge}$ by cross-correlating galaxies with the 2-d field $\bar\Delta(\th)$
of DM averaged over all FRB's detected in a particular direction $\th$.
Such statistics are related to the DM moment of $C_l^{fg}$:
\be
C_l^{\bar\Delta g} \propto \sum_i D_i n_{f_i}^{2d} C_l^{f_i g},
\ee
where $f_i$ denotes the sample of FRB's in DM bin $i$ centered on $D_i$. Since
$C_l^{\bar\Delta g}$ is a moment of our clustering statistic $C_l^{fg}$, the
former contains a subset of the astrophysical information.

\subsection{Numerical results}
\label{ssec:numerical_propagation_results}

In this section, we numerically compare contributions to $C_l^{fg}$ from spatial
clustering, and two propagation effects: DM-shifting (Eq.~(\ref{eq:dm_shifting}))
and completeness (Eq.~(\ref{eq:completeness})).
For the completeness effect, we have used FRB intrinsic width $t_i = 10^{-3}$ sec, 
and instrumental parameters matching CHIME/FRB: time sampling $t_s = 10^{-3}$ sec, 
channel bandwidth $\Delta\nu = 400$ kHz, and central frequency $\nu=600$ MHz.

To visualize contributions to $C_l^{fg}$, we compress the power spectrum into two
observables $b_f (dn_f^{2d}/dz)$ and $\gamma_f (dn_f^{2d}/dz)$, as described in~\S\ref{sec:power_spectrum_observables}.
To compute these observables for propagation effects, we split the galaxy-electron power spectrum $P_{ge}$ into 2-halo and 1-halo terms
(see Eq.~(\ref{eq:pge}) in Appendix~\ref{app:halo_model}).
In the limit of low-$l$, these take the forms
\ba
P_{ge}^{2h}(k,z) & \rightarrow & b_g(z) b_e(z) P_{\rm lin}(k,z) \label{eq:pge_2h_lowl} \\
P_{ge}^{1h}(k,z) & \rightarrow & \frac{n_{ge}^{3d}(z)}{n_g^{3d}(z) n_e^{3d}(z)}  \label{eq:pge_1h_lowl}
\ea
where $n_e^{3d}(z)$ is the 3-d number density of free electrons, and $n_{ge}^{3d}(z)$
is defined by:
\be
n_{ge}^{3d}(z) = \int dM \, n_h^{3d}(M,z) \, \big\langle N_g N_e \rangle_{M,z}  \label{eq:nge_def}
\ee
similar to the definition of $n_{fg}^{3d}(z)$ in Eq.~(\ref{eq:nfg_def}).
Now a calculation combining
Eqs.~(\ref{eq:clfg_2h_b}),~(\ref{eq:clfg_1h_gamma})~(\ref{eq:clfg_p_zbin}),~(\ref{eq:pge_2h_lowl}),~(\ref{eq:pge_1h_lowl})
shows that the contribution to the power spectrum observables $(b_f dn_f^{2d}/dz)$ and
$(\gamma_f dn_f^{2d}/dz)$ from propagation effects is:
\ba
\left( b_f \frac{dn_f^{2d}}{dz} \right)_{\rm prop} &=& W_f(z) \big( b_e(z) n_f^{2d} \big)  \label{eq:b_prop} \\
\left( \gamma_f \frac{dn_f^{2d}}{dz} \right)_{\rm prop} &=& W_f(z) \big( \gamma_e(z) n_f^{2d} \big)  \label{eq:gamma_prop}
\ea
Here, $b_e(z)$ is the large-scale clustering bias of free electrons, which
we will take to be 1.
The quantity $\gamma_e(z)$ is defined by:
\be
\gamma_e(z) = \frac{n_g^{3d}(z)}{n_e^{3d}(z)} \frac{n_{ge}^{3d}(z)}{n_{gg}^{3d}(z)}  \label{eq:gammae_def}
\ee
similar to the definition of $\gamma_f(z)$ given previously.

\begin{figure*}
\centerline{\includegraphics[width=15cm]{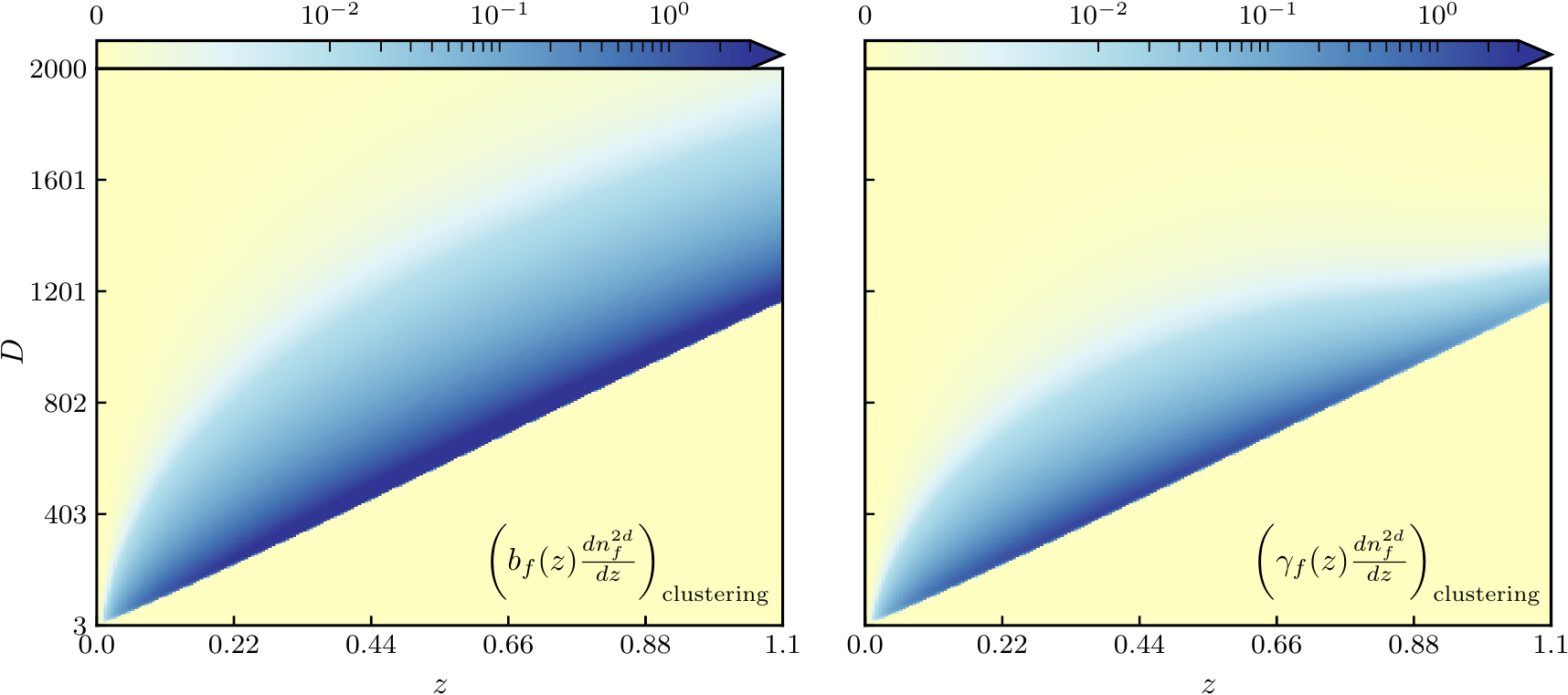}}
\vspace{-0.179cm}
\centerline{\includegraphics[width=15cm]{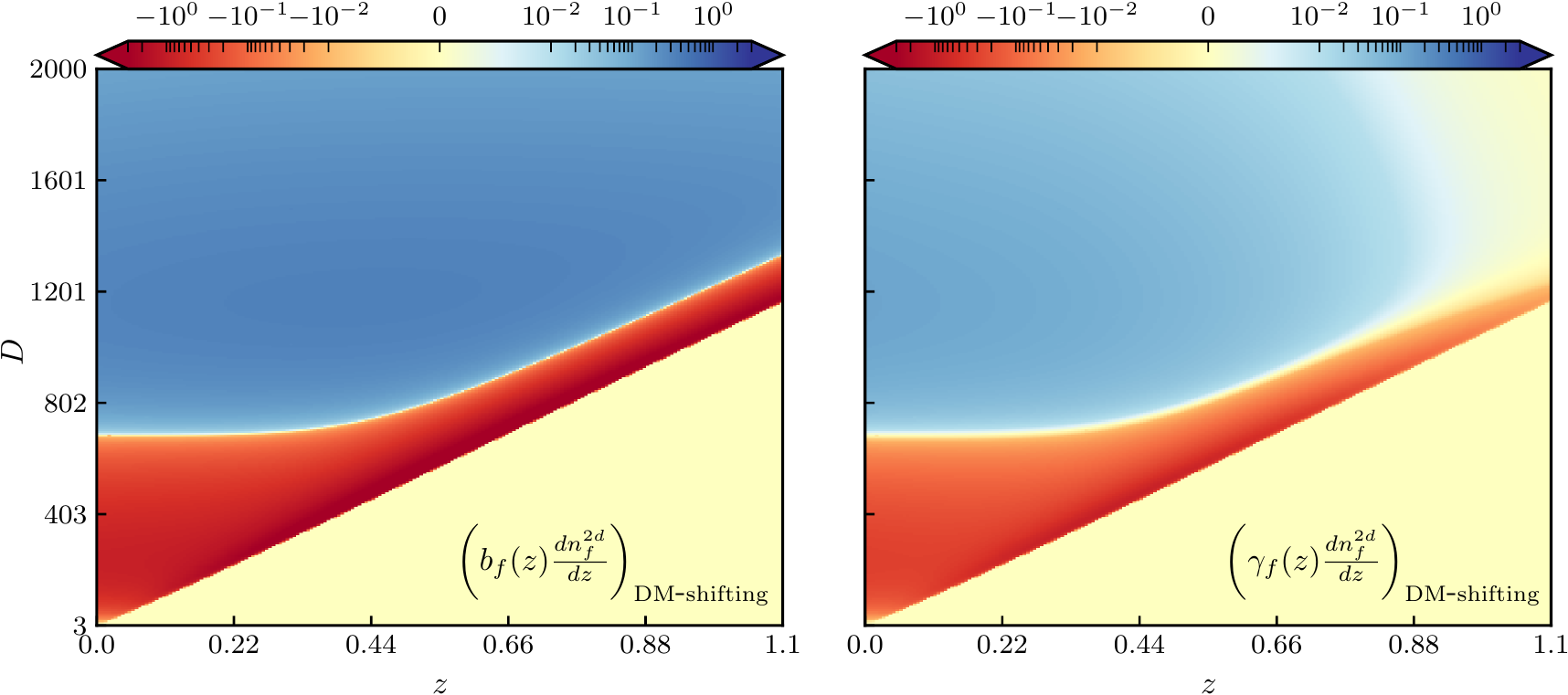}}
\vspace{-0.179cm}
\centerline{\includegraphics[width=15cm]{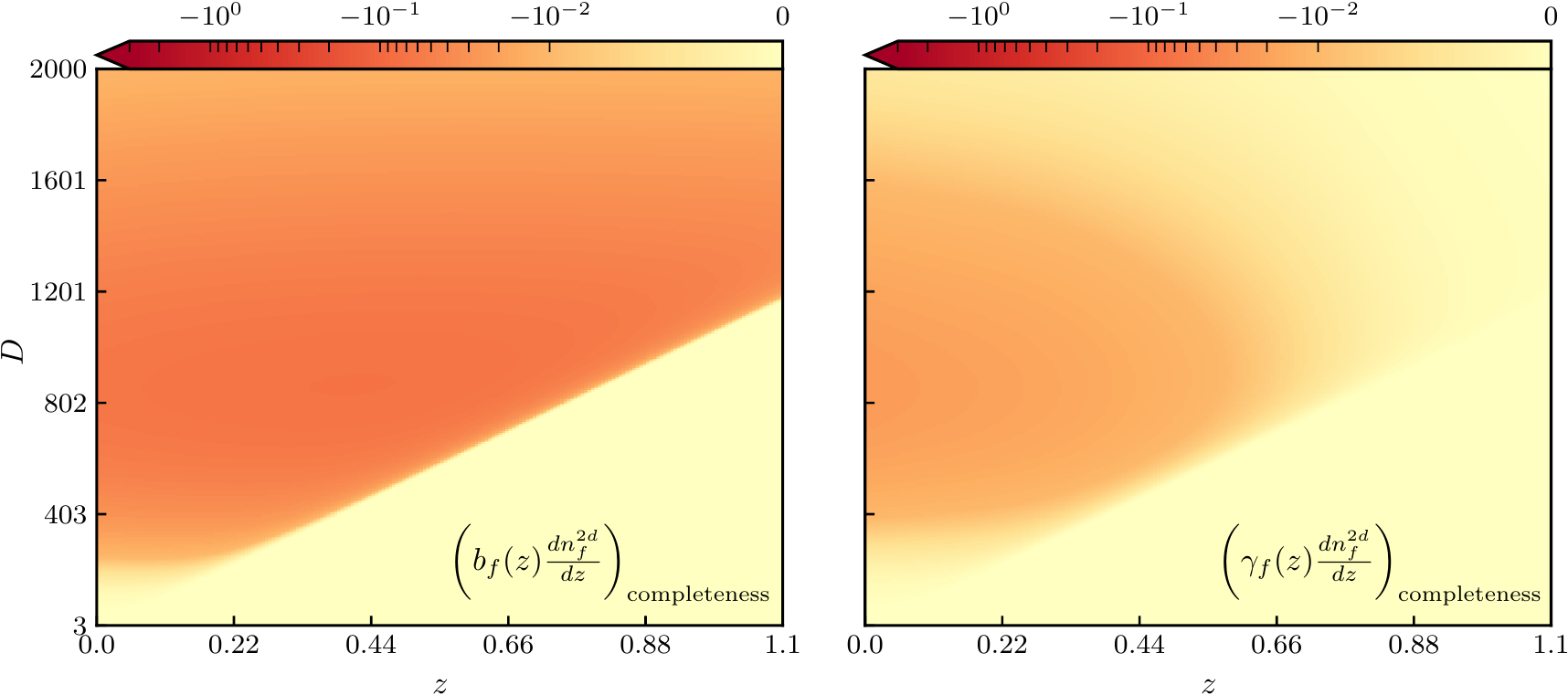}}
\caption{Visual comparison between clustering and propagation contributions to the clustering power spectrum $C_l^{fg}$,
for our high-$z$ fiducial FRB model and SDSS-DR8.
Each row corresponds to one such contribution: clustering (top), DM-shifting propagation effect (middle),
and completeness propagation effect (bottom).
Since $C_l^{fg}$ is a function of three variables $(z,D,l)$, we compress the $l$-dependence into two
clustering observables $b_f dn_f/dz$ (left column) and $\gamma_f dn_f/dz$ (right column), as described
in~\S\ref{sec:power_spectrum_observables}. 
Qualitatively, it is clear that clustering and propagation effects may be distinguished based on their $(z,D)$ dependence.}
\label{fig:obs_high_z}
\end{figure*}

\begin{figure*}
\centerline{\includegraphics[width=16cm]{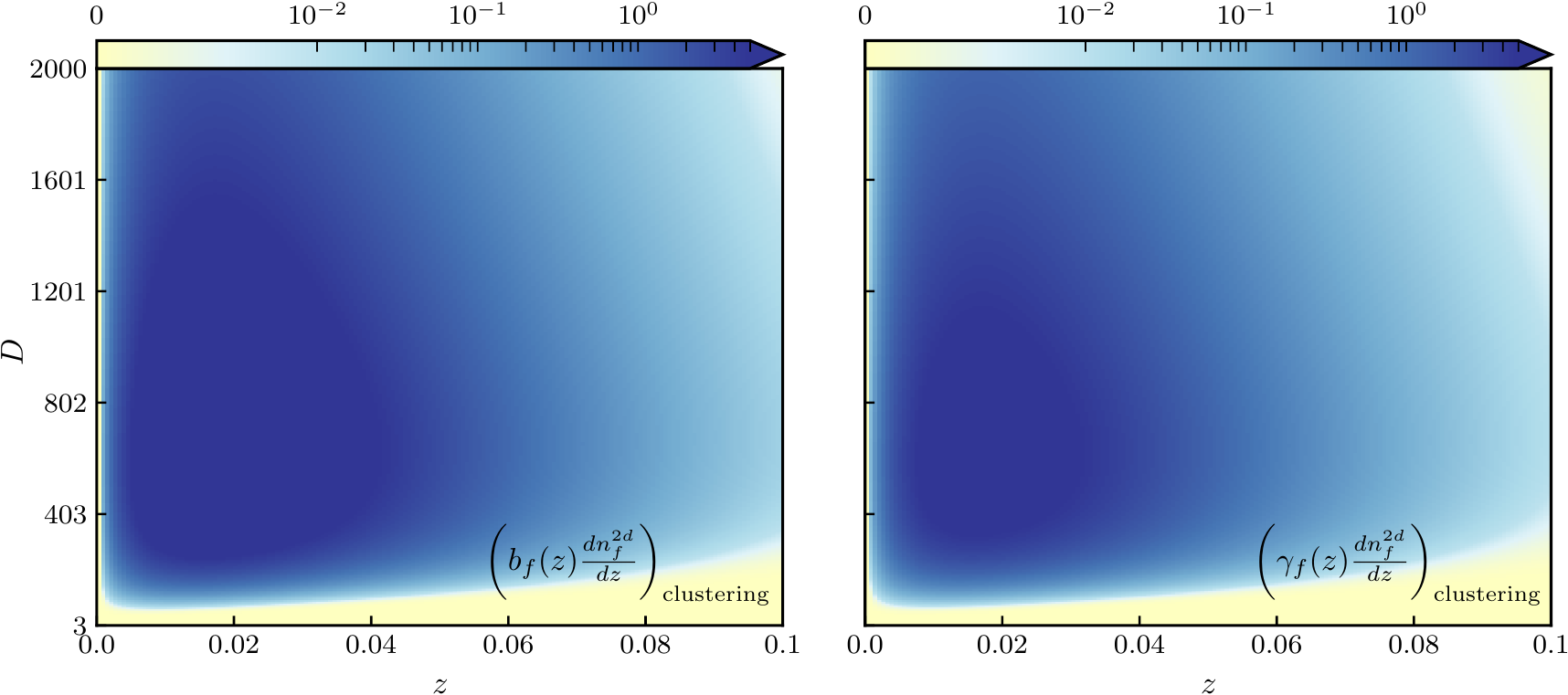}}
\vspace{-0.1875cm}
\centerline{\includegraphics[width=16cm]{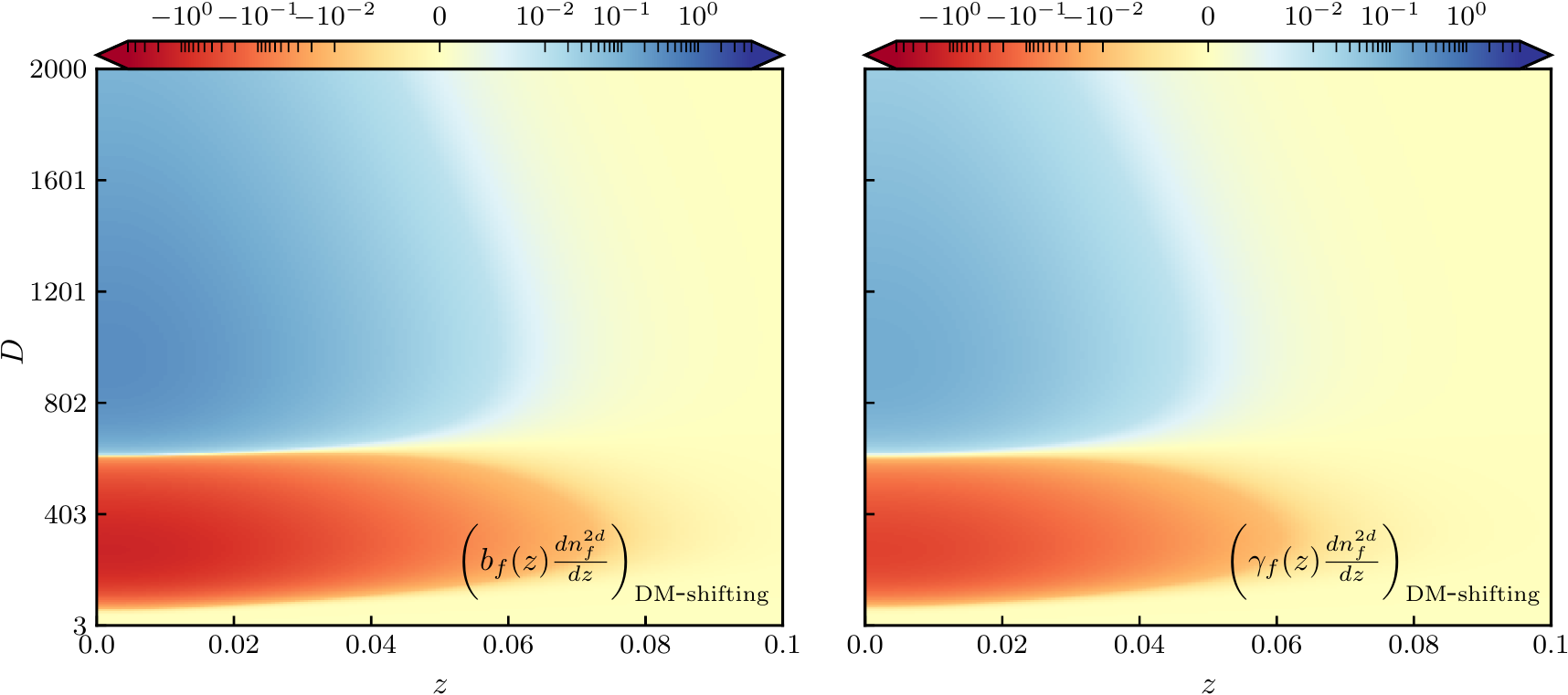}}
\vspace{-0.1875cm}
\centerline{\includegraphics[width=16cm]{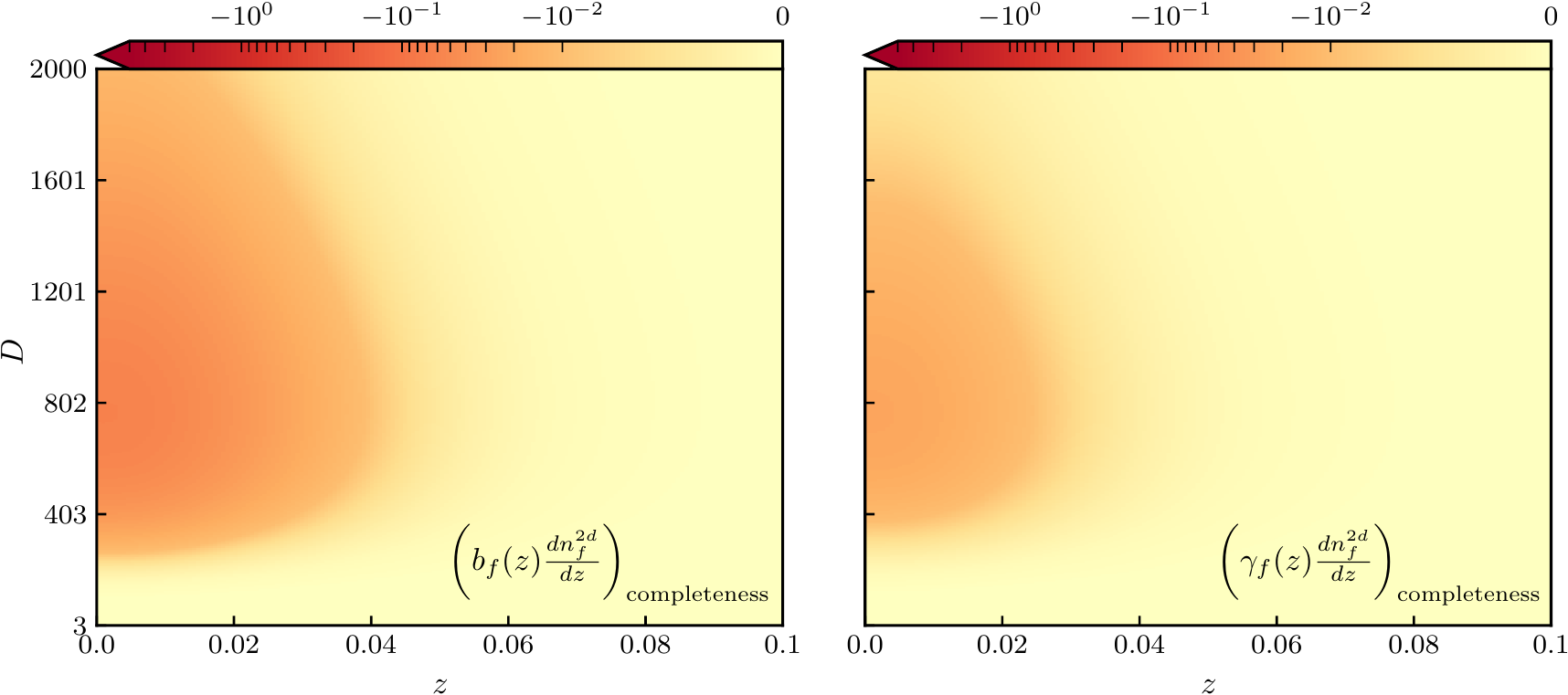}}
\caption{Same as Figure~\ref{fig:obs_high_z}, but for the low-$z$ fiducial FRB model.}
\label{fig:obs_low_z}
\end{figure*}

In Figures~\ref{fig:obs_high_z},~\ref{fig:obs_low_z}, we show power spectrum observables $b_f(dn_f^{2d}/dz)$ and $\gamma_f(dn_f^{2d}/dz)$
from clustering and both propagation effects, in the (DM, $z$) plane.
It is seen that propagation effects are comparable in size to the clustering signal!
However, it is qualitatively clear from Figures~\ref{fig:obs_high_z},~\ref{fig:obs_low_z}
that there is some scope for separating the two based on their dependence on redshift and DM.

\subsection{Ideas for separating spatial clustering from propagation effects}
\label{ssec:propagation_ideas}

Propagation effects complicate interpretation of the FRB-galaxy cross
spectrum $C_l^{fg}$.  For example, suppose a nonzero correlation is observed
between high-DM FRB's and low-redshift galaxies.  In the absence of propagation
effects, this would mean that the FRB's and galaxies must overlap
in redshift, implying a significant population of FRB's at low redshift and large
host DM.  However, in the presence of propagation effects, another possibility
is that FRB's are at high redshift, and correlated to low-redshift galaxies
via propagation effects.

On the other hand, propagation effects add new information to $C_l^{fg}$.
By treating propagation effects as signal rather than noise, it may be possible
to learn about the distribution of electrons in the IGM.
In this section, we will consider the question of how the spatial clustering and
propagation contributions to $C_l^{fg}$ might be separated.
Rather than trying to anticipate every observational scenario which may arise,
we will present some general ideas.

Propagation effects can sometimes be eliminated by changing the way the FRB catalog
is selected.  To take the case of dispersion, the DM-shifting term will be eliminated if
the FRB catalog is unbinned in DM.  Of course, this also throws away information since the
DM-dependence of the clustering signal is of interest.
The completeness term will be eliminated if FRB's are selected in a fluence bin,
rather than selecting FRB's above an SNR threshold.  The fluence bin must be complete,
in the sense that all FRB's in the bin are detected regardless of their dispersion.
This may require restricting the cross-correlation to fairly large fluence and discarding
low-fluence FRB's in the catalog.

Some propagation effects have a preferred sign, for example the completeness
term in Eq.~(\ref{eq:completeness}) is negative, since adding dispersion makes FRB's
harder to detect.\footnote{As discussed near Eq.~(\ref{eq:completeness}), this is true
for our toy instrumental model, but not guaranteed to be true for a real pipeline.}
Scattering is another example of a propagation effect with a negative sign, for the same reason.

Propagation effects appear in the $C_l^{fg}$ power spectrum
via the product $W_f(z) P_{ge}(l/\chi,z)$ (Eq.~(\ref{eq:clfg_p})).
We will discuss separately how the window function $W_f(z)$
and galaxy-electron power spectrum $P_{ge}(k,z)$ might be modeled.

The window function $W_f(z)$ may simplify in the limit of low $z$.
As a concrete example, consider the DM-shifting effect, where the window function is:
\ba
W_f(z) &=& n_{e,0} \frac{1+z}{H(z)} \\
&& \hspace{0.0cm} \times \int_z^{\infty} \frac{dz'}{n_f^{2d}}\left[
  \left( \frac{d^2n_f^{2d}}{dz' \, dD} \right)_{D_{\rm min}} \hspace{-0.3cm} - \hspace{0.1cm} \left( \frac{d^2n_f^{2d}}{dz' \, dD} \right)_{D_{\rm max}}\nn
\right]
\ea
by combining Eqs.~(\ref{eq:Wf_dispersion}),~(\ref{eq:dm_shifting}).
In the limit of low $z$ this becomes:
\be
\lim_{z\rightarrow 0} W_f(z) = \frac{n_{e,0}}{H_0} \frac{1}{n_f^{2d}} \left[
  \left( \frac{dn_f^{2d}}{dD} \right)_{D_{\rm min}} \hspace{-0.3cm} - \hspace{0.1cm} \left( \frac{dn_f^{2d}}{dD} \right)_{D_{\rm max}}
\right]
\ee
where the derivative $(dn_f^{2d}/dD)$ can be estimated directly from data, since
it is just the DM-derivative of the observed DM distribution.

A similar comment applies to other propagation effects: the $z \rightarrow 0$ limit of the
window function $W_f(z)$ can be estimated directly from the distribution of observed FRB parameters,
plus a model of the instrumental selection.
Away from the $z \rightarrow 0$ limit, the window function will depend on the FRB redshift
distribution, which is not directly observable.  On the other hand, this means that if the
$z$ dependence of $W_f(z)$ can be measured, it constrains the FRB redshift distribution.

Now we discuss modeling the galaxy-electron power spectrum $P_{ge}(k,z)$.
On 2-halo dominated scales, where $P_{ge}(k,z) = b_g(z) b_e(z) P_{\rm lin}(k,z)$,
this should be straightforward.
The galaxy bias $b_g(z)$ can be determined either from the galaxy auto power spectrum
or cross-correlations with gravitational lensing, and the electron bias $b_e(z)$ is
expected to be very close to 1.

On 1-halo dominated scales, modeling $P_{ge}(k,z)$ is more difficult.
One interesting near-future possibility is to measure $P_{ge}(k,z)$ through the kSZ
(kinetic Sunyaev-Zeldovich) effect in the cosmic microwave background.
Currently, the kSZ effect has been detected at a few sigma, but not constrained
to high precision.  However, measurements at the $\approx$10$\sigma$
level are imminent, and future CMB experiments such as Simons Observatory and
CMB-S4 will measure $P_{ge}$ with percent-level accuracy~\cite{Smith:2018bpn,Ade:2018sbj}.
These measurements will be very informative for modeling FRB propagation effects.

Less futuristically, the galaxy-{\em matter} power spectrum $P_{gm}(k,z)$ can be
measured using cross-correlations between the galaxy catalog and gravitational lensing maps.
On large scales, $P_{gm}(k,z)$ and $P_{ge}(k,z)$ are nearly equal, but on smaller scales they
will differ since dark matter halo profiles are expected to be more compact than electron profiles.
Nevertheless, measuring $P_{gm}$ may be a useful starting point for modeling $P_{ge}$.

In a scenario where $P_{ge}(k,z)$ has been measured accurately as a function of $k$,
the $l$-dependence of $C_l^{fg}$ is determined, even if the window function $W_f(z)$
is completely unknown.
Therefore, it is possible to marginalize over propagation effects by fitting and subtracting
a ($z$-dependent) multiple of $P_{ge}(l/\chi,z)$ from $C_l^{fg}$.
This marginalization will degrade clustering information to some extent.
In the two-observable picture, statistical errors would increase on one linear combination
of $b_f (dn_f^{2d}/dz)$ and $\gamma_f (dn_f^{2d}/dz)$.

Summarizing, there are several interesting ideas for modeling the separation
of $C_l^{fg}$ into clustering and propagation signals.  Which of these ideas
proves to be most useful will depend on which observational scenario emerges,
and what auxiliary information is available (e.g.~kSZ).

\section{Forecasts and signal-to-noise}
\label{sec:forecasts}

\subsection{Fisher matrix formalism}

Our basic forecasting tool is the Fisher matrix, which we briefly review.
Suppose we have $M$ FRB fields $f_1, \cdots, f_M$ and $N$ galaxy fields $g_1, \cdots, g_N$.
We will always assume that galaxy fields are defined by narrow redshift bins,
but FRB fields could be defined by binning in DM or a different quantity,
or the FRB field could be unbinned ($M=1$).

We assume the FRB-galaxy cross power spectrum is of the form:
\be
C_l^{f_ig_j} = \sum_\mu A_\mu C_l^{f_ig_j(\mu)}  \label{eq:fisher_fg}
\ee
where $\mu=1,\cdots,P$.  That is, the power spectrum is the sum of $P$
terms whose $l,i,j$ dependence is fixed by a model, but whose coefficients
$A_\mu$ are to be determined from data.
For example, we could take $\mu \in \{1h,2h\}$
with $P=2$, to forecast constraints on the overall amplitude of the 1-halo and
2-halo clustering terms.  Propagation effects can similarly be included in the
forecast.

Given this setup, the $P$-by-$P$ Fisher matrix is:
\be
F_{\mu\nu} = \Omega \sum_{ij} \int \frac{l\,dl}{2\pi} \frac{C_l^{f_ig_j(\mu)} C_l^{f_ig_j(\nu)}}{C_l^{f_if_i} C_l^{g_jg_j}}  \label{eq:fisher_matrix}
\ee
We assume that FRB auto power spectra are Poisson noise dominated, i.e.
\be
C_l^{f_if_i} = \big( n_{f_i}^{2d} \big)^{-1}
\ee
but have written $C_l^{f_if_i}$ in Eq.~(\ref{eq:fisher_matrix}) for notational uniformity.

The Fisher matrix is the forecasted {\em inverse} covariance matrix of the amplitude parameters $A_\mu$ in
Eq.~(\ref{eq:fisher_fg}).  For example, if $P=1$, then the 1-by-1 Fisher ``matrix'' $F$ is the SNR$^2$,
and the statistical error on the amplitude parameter is $\sigma(A) = 1/\sqrt{F}$.

A few technical comments.
The form of the Fisher matrix in Eq.~(\ref{eq:fisher_matrix}) assumes that FRB and galaxy fields are each uncorrelated, i.e.
\be
C_l^{f_if_j} = \delta_{ij} C_l^{f_if_i} \hspace{1cm} C_l^{g_ig_j} = \delta_{ij} C_l^{g_ig_i}
\ee
This assumption is satisfied for FRB fields, since we are assuming that auto spectra are Poisson noise
dominated.  The galaxy fields will also be uncorrelated if they are defined by a set of non-overlapping redshift bins.
Eq.~(\ref{eq:fisher_matrix}) also assumes that $C_l^{fg} \ll (C_l^{ff} C_l^{gg})^{1/2}$ in the fiducial model.
This will be a good approximation if the FRB number density is not too large.
Finally, in Eq.~(\ref{eq:fisher_matrix}) we have written the Fisher matrix as a double sum over (redshift, DM) bins
for maximum generality, but for numerical forecasts we take the limit of narrow bins, by replacing the sum by an appropriate
double integral.

\subsection{Numerical results}
\label{ssec:numerical_forecasts}

In Table~\ref{tab:snr}, we show SNR forecasts for several FRB and galaxy surveys.
We report SNR separately for six contributions to the power spectrum $C_l^{fg}$ as follows.
First, we split the power spectrum into three terms from gravitational clustering,
and the DM-shifting and completeness propagation effects described in~\S\ref{sec:propagation_effects}.
We then split each of these terms into 1-halo and 2-halo contributions, for a total of 6 terms.
Each SNR entry in Table~\ref{tab:snr} is given by $\sqrt{F_{ii}}$, where $F_{ii}$ is the appropriate
diagonal element of the 6-by-6 Fisher matrix.
This corresponds to SNR of each contribution considered individually, without marginalizing the amplitude
of the other terms in a joint fit.

\begin{table}
\begin{tabular}{|l|cc|cc|cc|} \hline
  & \multicolumn{2}{c|}{Clustering} &  \multicolumn{2}{c|}{DM-shifting} &  \multicolumn{2}{c|}{Completeness} \\
      & $1h$ & $2h$ & $1h$ & $2h$ & $1h$ & $2h$ \\
\hline
High-$z$ FRB model & & & & & &  \\
\hspace{0.1cm}SDSS-DR8, $\theta_f=1'$ & 25 & 6.1 & 18 & 5.8 & 1.2 & 0.4 \\ 
\hspace{0.1cm}SDSS-DR8, $\theta_f=10'$ & 6.9 & 5.8 & 8.3 & 5.6 & 0.57 & 0.38 \\ 
\hspace{0.1cm}SDSS-DR8, $\theta_f=30'$ & 2.4 & 4.9 & 5 & 4.9 & 0.34 & 0.33 \\ 
\hspace{0.1cm}2MPZ, $\theta_f=1'$ & 8.2 & 1.8 & 10 & 2.8 & 0.72 & 0.2 \\ 
\hspace{0.1cm}2MPZ, $\theta_f=10'$ & 4.8 & 1.7 & 7.4 & 2.8 & 0.51 & 0.2 \\ 
\hspace{0.1cm}2MPZ, $\theta_f=30'$ & 2.2 & 1.7 & 4.8 & 2.8 & 0.32 & 0.19 \\ 
\hspace{0.1cm}DESI-ELG, $\theta_f=1'$ & 12 & 4.6 & 5.4 & 3.4 & 0.34 & 0.22 \\ 
\hspace{0.1cm}DESI-ELG, $\theta_f=10'$ & 1.9 & 4.2 & 0.85 & 3.1 & 0.055 & 0.2 \\ 
\hspace{0.1cm}DESI-ELG, $\theta_f=30'$ & 0.49 & 3.2 & 0.22 & 2.4 & 0.014 & 0.15 \\ 
\hline
Low-$z$ FRB model & & & & & &  \\
\hspace{0.1cm}SDSS-DR8, $\theta_f=1'$ & 103 & 14 & 4.4 & 0.74 & 0.28 & 0.049 \\ 
\hspace{0.1cm}SDSS-DR8, $\theta_f=10'$ & 87 & 14 & 4.1 & 0.74 & 0.26 & 0.049 \\ 
\hspace{0.1cm}SDSS-DR8, $\theta_f=30'$ & 63 & 14 & 3.5 & 0.74 & 0.22 & 0.048 \\ 
\hspace{0.1cm}2MPZ, $\theta_f=1'$ & 92 & 13 & 3.9 & 0.7 & 0.25 & 0.046 \\ 
\hspace{0.1cm}2MPZ, $\theta_f=10'$ & 82 & 13 & 3.7 & 0.7 & 0.24 & 0.046 \\ 
\hspace{0.1cm}2MPZ, $\theta_f=30'$ & 62 & 13 & 3.2 & 0.7 & 0.21 & 0.046 \\ 
\hline
\end{tabular}
\caption{Forecasted SNR for FRB-galaxy cross-correlations.
 Each row corresponds to a choice of FRB model, galaxy survey,
 and FRB angular resolution $\theta_f$.
 Each column corresponds to one contribution to the FRB-galaxy power spectrum.
 Each entry is total SNR after summing over angular wavenumber $l$, and a narrow set of redshift and DM bins.
 We have assumed a catalog with $N_{\rm FRB}=1000$ FRB's ($D_{\rm max}=10^4$); in general each SNR
 value scales as $N_{\rm FRB}^{1/2}$.}
\label{tab:snr}
\end{table}

The forecasts are extremely promising: a CHIME/FRB-like experiment which measures
catalogs of $\sim 10^3$ FRB's with few-arcminute angular resolution can measure the clustering
signal with high SNR.
The precise value depends on the FRB redshift distribution and choice of galaxy survey, but
can be as large as $\approx 100$ in the low-$z$ FRB model.
As a consequence of the high total SNR, the FRB-galaxy correlation can be split up and
measured in $(z,D)$ bins, allowing the redshift distribution (or rather, the observables
$b_f dn_f/dz$ and $\gamma_f dn_f/dz$) to be measured.

One interesting feature of Table~\ref{tab:snr} is that if FRB's do extend to high redshift,
the cross-correlation with a high-redshift galaxy sample is detectable (e.g.~SNR=12 for
the high-$z$ FRB model, DESI-ELG, and $\theta_f=1$ arcminute).
Angular cross-correlations should be a powerful tool for probing the high-$z$ end of the
FRB redshift distribution, where galaxy surveys are far from complete, and FRB host galaxy
associations are difficult.

To get a sense for the level of correlation between different contributions to the FRB-galaxy
power spectrum, we rescale the Fisher matrix to a correlation matrix $r_{ij} = F_{ij} / (F_{ii} F_{jj})^{1/2}$
whose entries are between $-1$ and $1$.  In the high-$z$ FRB model, we get:
\be
\left( \begin{array}{cccccc}
     1.00 &  0.20 & -0.76 & -0.17 & -0.10 & -0.03 \\
     0.20 &  1.00 & -0.14 & -0.83 & -0.02 & -0.14 \\
    -0.76 & -0.14 &  1.00 &  0.19 & -0.22 & -0.04 \\
    -0.17 & -0.83 &  0.19 &  1.00 & -0.04 & -0.23 \\
    -0.10 & -0.02 & -0.22 & -0.04 &  1.00 &  0.19 \\
    -0.03 & -0.14 & -0.04 & -0.23 &  0.19 &  1.00
\end{array} \right)
\ee
where the row ordering is the same as Table~\ref{tab:snr}.
We see that there is not much correlation between 1-halo and 2-halo signals, but
the clustering signal is fairly anti-correlated to the DM-shifting signal.
The correlation is not perfect, since there is some difference in the (redshift, DM)
dependence, as can be seen directly by comparing the top and middle rows of Figure~\ref{fig:obs_high_z}.
The correlation matrix depends to some degree on model assumptions.
For example, in the low-$z$ FRB model, the correlation matrix is:
\be
\left( \begin{array}{cccccc}
    1.00 &  0.17 & -0.02 & -0.00 & -0.78 & -0.16 \\
    0.17 &  1.00 & -0.00 & -0.02 & -0.13 & -0.86 \\
   -0.02 & -0.00 &  1.00 &  0.19 & -0.19 & -0.04 \\
   -0.00 & -0.02 &  0.19 &  1.00 & -0.04 & -0.20 \\
   -0.78 & -0.13 & -0.19 & -0.04 &  1.00 &  0.19 \\
   -0.16 & -0.86 & -0.04 & -0.20 &  0.19 &  1.00
\end{array} \right)
\ee
Here, there is a large correlation between clustering and completeness terms.
(However, Table~\ref{tab:snr} shows that completeness terms are small in the low-$z$ FRB model.)

\begin{figure*}
\centerline{
  \includegraphics[width=8.6cm]{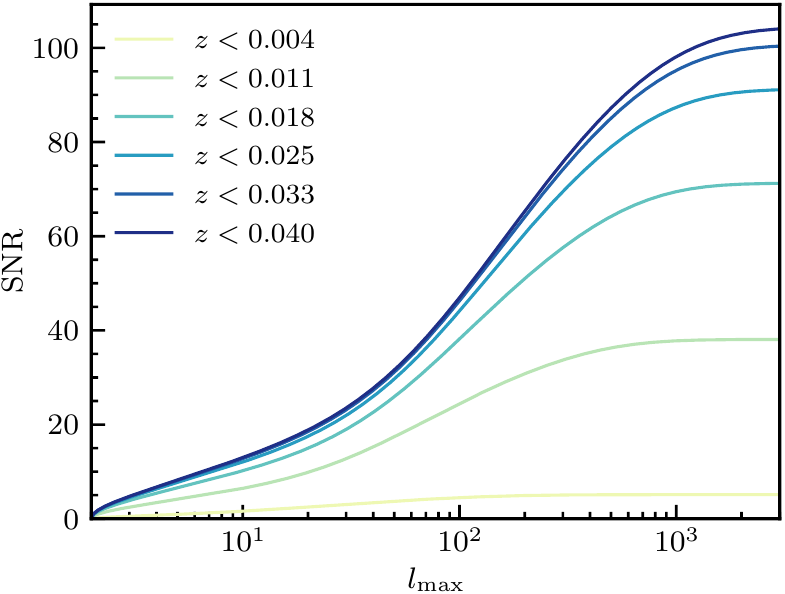}
  \hspace{0.35cm}
  \includegraphics[width=8.6cm]{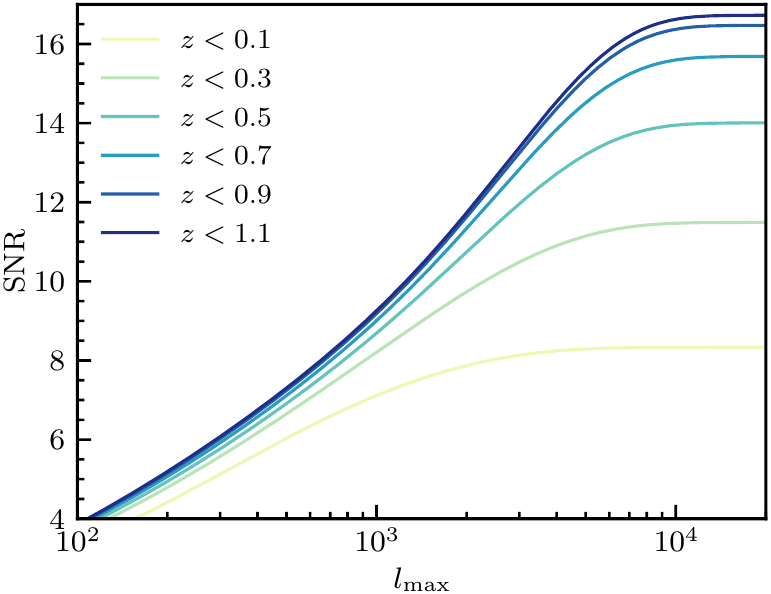}
}
\caption{Forecasted SNR of FRB-galaxy cross power, for varying choices of maximum redshift $z_{\rm max}$
and maximum angular multipole $l_{\rm max}$, after summing over narrow $(D,z)$ bins.
{\em Left panel:} Low-$z$ FRB fiducial model and SDSS-DR8 galaxies.
{\em Right panel:} High-$z$ FRB fiducial model and SDSS-DR8 galaxies.}
\label{fig:snr_forecast}
\end{figure*}

Figure~\ref{fig:snr_forecast} shows the evolution of total SNR as a function of angular
wavenumber and redshift.
In the analysis of real data, large scales ($l \lsim 20$) may be contaminated by Galactic
systematic effects, such as dust extinction.
Figure~\ref{fig:snr_forecast} shows that these scales make a small contribution to the total SNR,
so our forecasts are robust against such systematics.

\section{Simulations}
\label{sec:simulations}

Our SNR forecasts in the previous section make the approximation that the FRB and galaxy fields
are Gaussian.  More precisely, we are assuming that the bandpower covariance of the FRB-galaxy
power spectrum is given by the Gaussian (or disconnected) form:
\be
\Cov(C_b^{fg}, C_{b'}^{fg}) \approx \frac{C_b^{ff} C_b^{gg}}{f_{\rm sky} (l_{\rm max}^{(b)2} - l_{\rm min}^{(b)2})} \delta_{bb'}  \label{eq:bp_cov_gaussian}
\ee
where $C_b^{fg}$ denotes the estimated FRB-galaxy power in a set of non-overlapping
$l$-bands $l_{\rm min}^{(b)} \le l \le l_{\rm max}^{(b)}$ with $b=1,\cdots, N_{\rm bands}$,
and we have assumed $C_l^{fg} \ll (C_l^{ff} C_l^{gg})^{1/2}$.

In reality, FRB and galaxy fields are non-Gaussian.
The FRB catalog consists of a modest number of objects which obey Poisson (not Gaussian) statistics.
Galaxy catalogs are larger, but Poisson statistics of the underlying halos
may be important, since the number of halos is smaller than the number of galaxies.
The purpose of this section is to determine whether the Gaussian covariance~(\ref{eq:bp_cov_gaussian})
is a good approximation, by carrying out Monte Carlo simulations of galaxies and FRB's.

\subsection{Description of simulation pipeline}

Our simulation pipeline is based on the halo model from~\S\ref{sec:preliminaries} and Appendix~\ref{app:halo_model}.
We use the high-$z$ FRB model.
Because non-Gaussian effects are expected to be largest for the 1-halo term, our simulation pipeline only includes
1-halo clustering.  In particular, we do not simulate the Gaussian linear density field $\delta_{\rm lin}$,
because it is not needed to simulate 1-halo clustering.

We use a $10\times10$~deg$^2$~sky patch, in the flat-sky approximation with periodic boundary conditions.
We sample Poisson random halos in 100 redshift bins, and 500 logarithmically-spaced mass bins between $M_f$
and $M_{\rm max}=10^{17}h^{-1} M_\odot$.
For each halo, we assign an FRB and galaxy count by sampling a Poisson random variable whose
expectation value is given by the HOD's in Eqs.~(\ref{eq:gal_hod}),~(\ref{eq:frb_hod}).
For each FRB and galaxy, we assign a 3-d location within the halo using the NFW profile~(Eq.~(\ref{eq:nfw_real})).
Angular positions are computed by projecting 3-d positions onto the sky patch.
In the case of FRB's, we convolve sky locations by the beam~(Eq.~(\ref{eq:bl_def})).
Finally, FRB's are assigned a random DM, which is the sum of the IGM contribution $D_i(z)$
and a random host contribution $D_h$ (see Eq.~(\ref{eq:dh_pdf})).

Next, we grid the FRB and galaxy catalogs onto a real-space $2049\times2049$
pixelization with resolution $\approx 0.3$ arcmin,
using the cloud-in-cell (CIC) weighting scheme.
We take the Fourier transform to obtain Fourier-space fields $\delta_f(\l)$, $\delta_g(\l)$.
Then, following Eq.~(\ref{eq:clxy_double_sum}), we estimate the angular cross power spectrum $C_l^{fg}$
by averaging the cross power $\langle \delta_f(\l)^* \, \delta_g(\l) \rangle$ in a non-overlapping
set of $l$-bins.

\subsection{Numerical results}
\begin{figure}
\includegraphics[width=8.6cm]{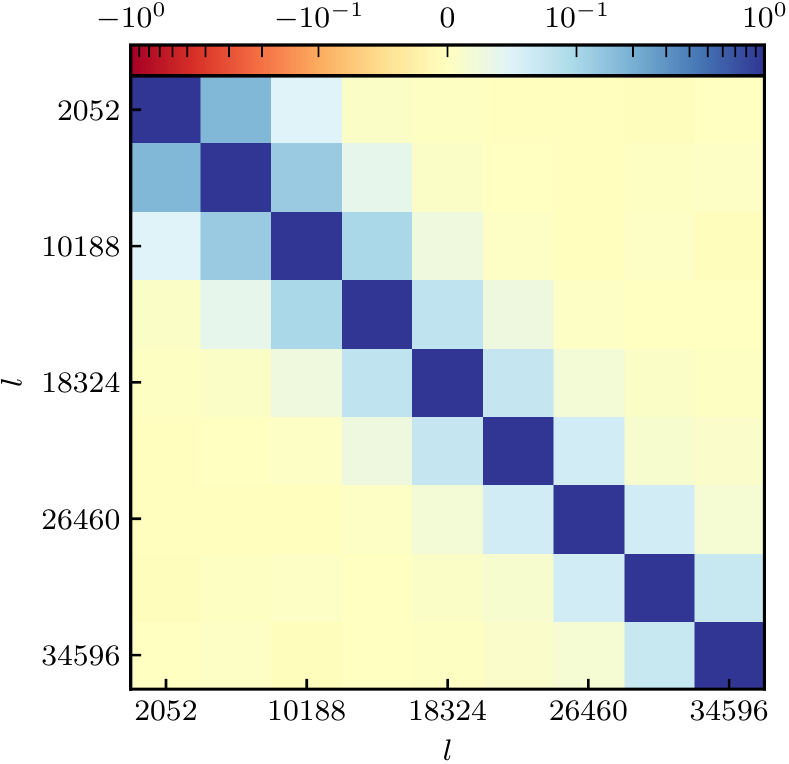}
\caption{Bandpower correlation matrix $r_{bb'}$ of the FRB-galaxy cross power spectrum $C_l^{fg(1h)}$, estimated from
simulations (see Eq.~(\ref{eq:bp_corr})).  We have used the high-$z$ fiducial FRB model, SDSS-DR8 galaxies,
FRB angular resolution $\theta_f = 1'$, and maximum dispersion measure $D_{\rm max}=10^4$.
Correlations between bandpowers are $\approx20\%$ for adjacent $l$-bins, and decay rapidly after that.
This is one way of quantifying the importance of non-Gaussian statistics, since off-diagonal correlations
would be zero if the FRB and galaxy fields were Gaussian.}
\label{fig:sim_corr}
\end{figure}

We run the pipeline for $10^5$ MC realizations and find that the cross power spectrum
$C_l^{fg}$ of the simulations agrees with the numerical calculation of $C_l^{fg(1h)}$,
for a few (DM,$z$) binning schemes.
To compare the bandpower {\em covariance} to the Gaussian approximation in Eq.~(\ref{eq:bp_cov_gaussian}),
we first estimate the covariance of the simulations as:
\ba
\Cov(C_b^{fg}, C_{b'}^{fg}) &=& \left(\frac{1}{n_{\rm sim}-1}\right) \label{eq:bp_cov_sim}  \\
&& \hspace{0.0cm} \times \sum_{i=1}^{n_{\rm sim}}
(C_{b}^{fg,i}-\langle C_{b}^{fg}\rangle)(C_{b^{\prime}}^{fg,i}-\langle
C_{b^{\prime}}^{fg}\rangle) \nn
\ea
In Figure~\ref{fig:sim_corr}, we show the bandpower {\em correlation} matrix $r_{bb'}$, obtained
from the Monte Carlo covariance matrix $C_{bb'}$ in Eq.~(\ref{eq:bp_cov_gaussian}) by
\be
r_{bb'} = \frac{C_{bb'}}{(C_{bb} C_{b'b'})^{1/2}}  \label{eq:bp_corr}
\ee
For a Gaussian field, $r_{bb'}$ is the identity (distinct bandpowers are uncorrelated).
In our simulations, we do see off-diagonal correlations due to non-Gaussian statistics,
but the correlations are small ($\approx$20\% for adjacent bands).

\begin{figure}
\includegraphics[width=8.6cm]{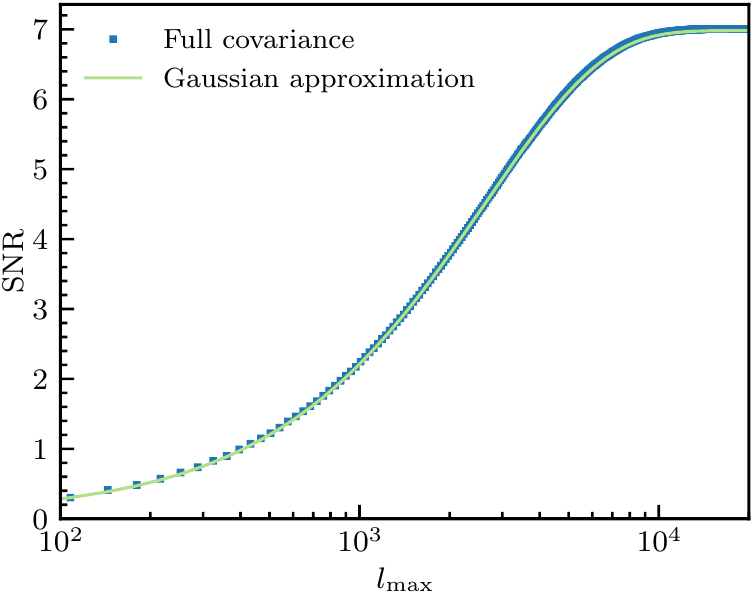}
\caption{Cumulative SNR for the FRB-galaxy cross power spectrum $C_l^{fg(1h)}$,
using the Monte Carlo bandpower covariance (Eq.~(\ref{eq:bp_cov_sim})),
with the Gaussian approximation shown for comparison (Eq.~(\ref{eq:bp_cov_gaussian})).
The two agree almost perfectly, justifying the Gaussian forecasts used throughout the paper.
We have used the high-$z$ fiducial FRB model, SDSS-DR8 galaxies, FRB angular resolution $\theta_f = 1'$,
and maximum dispersion measure $D_{\rm max}=10^4$.}
\label{fig:sim_snr}
\end{figure}

In Figure~\ref{fig:sim_snr}, we compare the total SNR of the FRB-galaxy cross correlation
obtained from simulations to the Gaussian approximation.  The total SNR was computed as:
\be
{\rm SNR}^2=\sum_{b,b'} (C_{b}^{fg}) \, \Cov(C_b^{fg}, C_{b'}^{fg})^{-1} (C_{b'}^{fg})  \label{eq:sim_snr}
\ee
where $\Cov(C_b^{fg}, C_{b'}^{fg})$ is either the Monte Carlo covariance matrix in
Eq.~(\ref{eq:bp_cov_sim}) or the Gaussian approximation in Eq.~(\ref{eq:bp_cov_gaussian}).
From Figure~\ref{fig:sim_snr}, the total SNR in the simulations agrees almost perfectly with
the Gaussian forecast.
This indicates that our forecasts in previous sections, which assume Gaussian statistics,
are good approximations to the true non-Gaussian statistics of the FRB and galaxy fields.

\section{Discussion}
\label{sec:discussion}

In summary, use of angular cross-correlations allows telescopes with high mapping
speed and modest angular resolution to constrain quantities which appear to require
host galaxy associations, such as the FRB redshift distribution.
Angular cross-correlations may also be detectable at high redshift, where galaxy
surveys are far from complete, and FRB host galaxy associations are difficult.
This dramatically extends the scientific reach of instruments like CHIME/FRB.

One complication is that the FRB redshift distribution $(dn_f/dz)$ is not quite
directly measurable.  In~\S\ref{sec:power_spectrum_observables} we studied this issue and
showed that there are two clustering observables $(b_f dn_f/dz)$ and $(\gamma_f dn_f/dz)$
in the 2-halo and 1-halo regimes respectively.  Here, $b_f$ is the usual large-scale bias
parameter, and the quantity $\gamma_f$ (defined in Eq.~(\ref{eq:gamma_def})) depends on
details of HOD's.

Propagation effects can produce
contributions to $C_l^{fg}$ which are comparable to the intrinsic clustering signal.
This means, for example, that if a nonzero correlation is observed between FRB's
and low-redshift galaxies, one cannot definitively conclude that a substantial population
of FRB's exists at low $z$.
The correlation could instead be due to the clustering of low-$z$ galaxies with free
electrons, which modulate the abundance of FRB's observed at higher $z$ through
either selection effects or by shifting FRB's between DM bins.

Propagation effects can be separated from clustering based on their dependence as
functions of $(z,D,l)$.  This is shown qualitatively in Figures~\ref{fig:obs_high_z} and~\ref{fig:obs_low_z},
where clustering and propagation signals have very different $(z,D)$ dependence
(after compressing the $l$ dependence into the two clustering observables $(b_f dn_f/dz)$
and $(\gamma_f dn_f/dz)$).
For a longer, more systematic discussion, see~\S\ref{ssec:propagation_ideas}.

Propagation effects are both a potential contaminant of the clustering signal,
and a potential source of information about ionized electrons in the universe.
Indeed, the ``DM-shifting'' propagation effect identified in~\S\ref{sec:propagation_effects}
can be used to probe the distribution of electrons in the CGM, along the lines
of~\cite{McQuinn:2013tmc,Masui:2015ola,Shirasaki:2017otr,Ravi:2018ose,Munoz:2018mll,Madhavacheril:2019buy}.

We now interpret our forecasts in relation to the 3$\sigma$ correlation between
ASKAP-discovered FRB's and 2MPZ galaxies measured in \cite{Li:2019fsg}. Scaling
to a sample of 21 galaxies, and noting the weak dependence on angular
resolution, our low-$z$ FRB model predicts an intrinsic clustering correlation SNR of roughly 12, a
factor of 4 higher than that observed. While it is not straightforward to
interpret SNR units---the difference could be one of either signal
amplitude, estimator optimally, or modeling---this would nonetheless seem to disfavor a
completely nearby population. However, the measured SNR is far greater than what
our high-$z$ FRB model predicts and cannot be explained by DM-shifting (the measurement was
unbinned in DM) or completeness as modeled (wrong sign and too small of an
amplitude).  As such, we suggest that the true FRB population may be somewhere
between these two models, which could still be consistent with the 3 direct
localizations (at high redshifts $z=0.19$, 0.32, 0.66).

The results in this paper can be extended in several directions.
We have not considered all possible propagation effects (e.g.~scattering, plasma lensing),
or fully explored the impact of various model assumptions (e.g.~free electron profiles).
We have explored the effect of binning the FRB catalog in DM, but not binning in other FRB observables.
One particularly interesting possibility will be binning FRB's by observed flux $F$.
By measuring the FRB distribution $d^2n_f/(dz\,dF)$ as a function of redshift and flux,
the intrinsic luminosities of FRB's can be constrained.

The galaxy catalog can also be binned in different ways.
As an interesting example which also illustrates subtleties in the interpretation,
suppose we bin galaxies by estimated star formation rate, in order to determine
whether FRB's are statistically associated with star formation.
If the FRB-galaxy correlation is observed to be larger for star-forming galaxies,
how should this be interpreted?

The answer depends on the angular scale $l$ where the power spectrum $C_l^{fg}$
is measured.  On angular scales which are 2-halo dominated, FRB's and galaxies
correlate via the observable $(b_f b_g dn_f/dz)$, so the observation just means
that the galaxy bias $b_g$ is larger for star-forming galaxies.
On 1-halo dominated scales, the observation would imply that FRB's preferentially
inhabit {\em halos} which contain star-forming galaxies, but this does not necessarily
imply that FRB's inhabit the star-forming galaxies themselves.
Finally, at very high $l$ where $C_l^{fg}$ is dominated by the Poisson term
(a regime which we have mostly ignored in this paper, but see discussion in~\S\ref{sec:clfg}),
the observation would imply that FRB's do preferentially inhabit star-forming galaxies.

In this paper, we have developed tools for analysis and interpretation
of FRB-galaxy cross correlations.  This work was largely motivated by
analysis of CHIME/FRB data in progress, to be reported separately in the
near future.

\vskip 0.2cm

{\em Acknowledgements.}
KMS was supported by an NSERC Discovery Grant, an Ontario Early Researcher Award, and
a CIFAR fellowship. Research at Perimeter Institute is supported in part by the Government of
Canada through the Department of Innovation, Science and Economic Development Canada and
by the Province of Ontario through the Ministry of Economic Development, Job Creation and Trade.
We thank Utkarsh Giri, Vicky Kaspi, Dustin Lang, Dongzi Li, and Ue-Li Pen for discussions.

\bibliographystyle{prsty}
\bibliographystyle{h-physrev}
\bibliography{frbx_paper}

\begin{thebibliography}{10}

\bibitem{Lorimer:2007qn}
D.~R. Lorimer, M.~Bailes, M.~A. McLaughlin, D.~J. Narkevic, and F.~Crawford,
\newblock Science {\bf 318}, 777 (2007), 0709.4301.

\bibitem{Katz:2016dti}
J.~I. Katz,
\newblock Mod. Phys. Lett. {\bf A31}, 1630013 (2016), 1604.01799.

\bibitem{Platts:2018hiy}
E.~Platts {\em et~al.},
\newblock (2018), 1810.05836.

\bibitem{Petroff:2019tty}
E.~Petroff, J.~W.~T. Hessels, and D.~R. Lorimer,
\newblock Astron. Astrophys. Rev. {\bf 27}, 4 (2019), 1904.07947.

\bibitem{Cordes:2002wz}
J.~M. Cordes and T.~J.~W. Lazio,
\newblock (2002), astro-ph/0207156.

\bibitem{YKW}
J.~M.~Y. Yao, R.~N. Manchester, and N.~Wang,
\newblock (2016), 1610.09448.

\bibitem{Petroff:2016tcr}
E.~Petroff {\em et~al.},
\newblock Publ. Astron. Soc. Austral. {\bf 33}, e045 (2016), 1601.03547.

\bibitem{Spitler:2016dmz}
L.~G. Spitler {\em et~al.},
\newblock Nature {\bf 531}, 202 (2016), 1603.00581.

\bibitem{Scholz:2016rpt}
P.~Scholz {\em et~al.},
\newblock Astrophys. J. {\bf 833}, 177 (2016), 1603.08880.

\bibitem{Amiri:2019bjk}
CHIME/FRB, M.~Amiri {\em et~al.},
\newblock Nature {\bf 566}, 235 (2019), 1901.04525.

\bibitem{Andersen:2019yex}
CHIME/FRB, B.~C. Andersen {\em et~al.},
\newblock Astrophys. J. {\bf 885}, L24 (2019), 1908.03507.

\bibitem{Chatterjee:2017dqg}
S.~Chatterjee {\em et~al.},
\newblock Nature {\bf 541}, 58 (2017), 1701.01098.

\bibitem{Marcote:2017wan}
B.~Marcote {\em et~al.},
\newblock Astrophys. J. {\bf 834}, L8 (2017), 1701.01099.

\bibitem{Tendulkar:2017vuq}
S.~P. Tendulkar {\em et~al.},
\newblock Astrophys. J. {\bf 834}, L7 (2017), 1701.01100.

\bibitem{Bannister:2019iju}
K.~W. Bannister {\em et~al.},
\newblock (2019), 1906.11476.

\bibitem{Ravi:2019alc}
V.~Ravi {\em et~al.},
\newblock Nature {\bf 572}, 352 (2019), 1907.01542.

\bibitem{Eftekhari:2017tbx}
T.~Eftekhari and E.~Berger,
\newblock Astrophys. J. {\bf 849}, 162 (2017), 1705.02998.

\bibitem{kiyo_beamforming}
K.~W. Masui {\em et~al.},
\newblock 1710.08591.

\bibitem{Amiri:2018qsq}
CHIME/FRB, M.~Amiri {\em et~al.},
\newblock (2018), 1803.11235.

\bibitem{Aihara:2011sj}
SDSS Collaboration, H.~Aihara {\em et~al.},
\newblock Astrophys. J. Suppl. {\bf 193}, 29 (2011), 1101.1559.

\bibitem{McQuinn:2013ib}
M.~McQuinn and M.~White,
\newblock Mon. Not. Roy. Astron. Soc. {\bf 433}, 2857 (2013), 1302.0857.

\bibitem{Menard:2013aaa}
B.~M\'enard {\em et~al.},
\newblock (2013), 1303.4722.

\bibitem{Rahman:2014lfa}
M.~Rahman, B.~M\'enard, R.~Scranton, S.~J. Schmidt, and C.~B. Morrison,
\newblock Mon. Not. Roy. Astron. Soc. {\bf 447}, 3500 (2015), 1407.7860.

\bibitem{Kovetz:2016hgp}
E.~D. Kovetz, A.~Raccanelli, and M.~Rahman,
\newblock Mon. Not. Roy. Astron. Soc. {\bf 468}, 3650 (2017), 1606.07434.

\bibitem{Passaglia:2017lnq}
S.~Passaglia, A.~Manzotti, and S.~Dodelson,
\newblock Phys. Rev. {\bf D95}, 123508 (2017), 1702.03004.

\bibitem{Hui:2007cu}
L.~Hui, E.~Gaztanaga, and M.~LoVerde,
\newblock Phys. Rev. {\bf D76}, 103502 (2007), 0706.1071.

\bibitem{Masui:2015ola}
K.~W. Masui and K.~Sigurdson,
\newblock Phys. Rev. Lett. {\bf 115}, 121301 (2015), 1506.01704.

\bibitem{McQuinn:2013tmc}
M.~McQuinn,
\newblock Astrophys. J. {\bf 780}, L33 (2014), 1309.4451.

\bibitem{Shirasaki:2017otr}
M.~Shirasaki, K.~Kashiyama, and N.~Yoshida,
\newblock Phys. Rev. {\bf D95}, 083012 (2017), 1702.07085.

\bibitem{Ravi:2018ose}
V.~Ravi,
\newblock Astrophys. J. {\bf 872}, 88 (2019), 1804.07291.

\bibitem{Munoz:2018mll}
J.~B. Muñoz and A.~Loeb,
\newblock Phys. Rev. {\bf D98}, 103518 (2018), 1809.04074.

\bibitem{Madhavacheril:2019buy}
M.~S. Madhavacheril, N.~Battaglia, K.~M. Smith, and J.~L. Sievers,
\newblock (2019), 1901.02418.

\bibitem{Li:2019fsg}
D.~Li, A.~Yalinewich, and P.~C. Breysse,
\newblock (2019), 1902.10120.

\bibitem{Bannister:2017sie}
K.~Bannister {\em et~al.},
\newblock Astrophys. J. {\bf 841}, L12 (2017), 1705.07581.

\bibitem{2018Natur.562..386S}
R.~M. {Shannon} {\em et~al.},
\newblock \nat {\bf 562}, 386 (2018).

\bibitem{Cooray:2002dia}
A.~Cooray and R.~K. Sheth,
\newblock Phys. Rept. {\bf 372}, 1 (2002), astro-ph/0206508.

\bibitem{Sheldon:2011fm}
E.~S. Sheldon, C.~Cunha, R.~Mandelbaum, J.~Brinkmann, and B.~A. Weaver,
\newblock Astrophys. J. Suppl. {\bf 201}, 32 (2012), 1109.5192.

\bibitem{Bilicki:2013sza}
M.~{Bilicki}, T.~H. {Jarrett}, J.~A. {Peacock}, M.~E. {Cluver}, and
  L.~{Steward},
\newblock Astrophys. J. Suppl. {\bf 210}, 9 (2014), 1311.5246.

\bibitem{Aghamousa:2016zmz}
DESI, A.~Aghamousa {\em et~al.},
\newblock (2016), 1611.00036.

\bibitem{Smith:2018bpn}
K.~M. Smith {\em et~al.},
\newblock (2018), 1810.13423.

\bibitem{Ade:2018sbj}
Simons Observatory, P.~Ade {\em et~al.},
\newblock JCAP {\bf 1902}, 056 (2019), 1808.07445.

\bibitem{Lewis:1999bs}
A.~Lewis, A.~Challinor, and A.~Lasenby,
\newblock \apj {\bf 538}, 473 (2000), astro-ph/9911177.

\bibitem{Sheth:2001dp}
R.~K. Sheth and G.~Tormen,
\newblock Mon. Not. Roy. Astron. Soc. {\bf 329}, 61 (2002), astro-ph/0105113.

\bibitem{Reed:2006rw}
D.~Reed, R.~Bower, C.~Frenk, A.~Jenkins, and T.~Theuns,
\newblock Mon. Not. Roy. Astron. Soc. {\bf 374}, 2 (2007), astro-ph/0607150.

\bibitem{Navarro:1996gj}
J.~F. Navarro, C.~S. Frenk, and S.~D.~M. White,
\newblock Astrophys. J. {\bf 490}, 493 (1997), astro-ph/9611107.

\bibitem{Dutton:2014xda}
A.~A. Dutton and A.~V. Macciò,
\newblock Mon. Not. Roy. Astron. Soc. {\bf 441}, 3359 (2014), 1402.7073.

\bibitem{Eke:1997ef}
V.~R. Eke, J.~F. Navarro, and C.~S. Frenk,
\newblock Astrophys. J. {\bf 503}, 569 (1998), astro-ph/9708070.

\bibitem{Battaglia:2016xbi}
N.~Battaglia,
\newblock JCAP {\bf 1608}, 058 (2016), 1607.02442.

\end{thebibliography}

\appendix

\section{Halo model}
\label{app:halo_model}

In this appendix, we describe the model for spatial clustering
of FRB and galaxies used throughout the paper.  We use a halo
model approach: first we specify the clustering of dark matter
halos, then specify how halos are populated by FRB's and galaxies.

\subsection{Dark matter halos}

We define $\sigma(R,z)$ to be the RMS amplitude of the {\em linear} density field
at redshift $z$, smoothed with a tophat filter of comoving radius $R$:
\be
\sigma(R,z) = \left( \int \frac{d^3k}{(2\pi)^3} P_{\rm lin}(k,z) W(kR)^2 \right)^{1/2}
\ee
where $W(x)$ is the Fourier transform of a unit-radius tophat:
\be
W(x) = 3 \, \frac{\sin(x) - x \cos(x)}{x^3}
\ee
and $P_{\rm lin}(k,z)$ is the matter power spectrum in linear perturbation theory,
which we compute numerically with CAMB~\cite{Lewis:1999bs}.  Throughout, we adopt a flat
$\Lambda$CDM cosmology with $h=0.67$, $\Omega_m=0.315$, $\Omega_b=0.048$,
$A_s=2\times10^{-9}$, $n_s=0.965$, $\sum_\nu m_\nu=0.06$~eV, and $T_{\rm CMB}=2.726$~K\@.

If $M$ is a halo mass, we define
\be
R_M = \left( \frac{3M}{4\pi\rho_m} \right)^{1/3}
\ee
where $\rho_m$ is the comoving total matter density (dark matter + baryonic).
Note that $R_M$ is just the radius of a sphere which encloses mass $M$ in a homogeneous universe.
Abusing notation slightly, we define $\sigma(M,z)$ to be equal to $\sigma(R,z)$ evaluated at $R=R_M$.

Let $n_h(M,z)$ be the halo mass function, i.e.~the number density of
halos per comoving volume per unit halo mass.
We use the Sheth-Tormen mass function~\cite{Sheth:2001dp,Reed:2006rw}, given by:
\ba
n_h(M) &=& \frac{\rho_{m,0}}{M} \frac{d \log \sigma^{-1}}{dM} f(\sigma) \label{eq:sheth_tormen_halo_mass_function}\\
f(\sigma) &=& A \frac{\delta_c}{\sigma} \sqrt{\frac{2a}{\pi}} \left( 1 + \left( \frac{\sigma^2}{a\delta_c^2} \right)^p \right) \exp\left( - \frac{a \delta_c^2}{2 \sigma^2} \right) \nn
\ea
where $\sigma=\sigma(M,z)$ and
\be
a = 0.707  \hspace{1.0cm}  \delta_c = 1.686  \hspace{1.0cm}   p = 0.3
\ee
and $A = 0.3222$ is the normalization which satisfies $\int d(\log \sigma) f(\sigma) = 1$,
which means that all matter is formally contained in halos of some (possibly very small) mass $M$.

We assume that halos are linearly biased Poisson tracers of the cosmological
linear density field $\delta_{\rm lin}$, i.e.~the number of halos in comoving
volume $V$ and mass range $(M,M+dM)$ is a Poisson random variable with mean
$dM (dn/dM) \int_V d^3x \, (1 + b_h(M) \delta_{\rm lin}(x))$.
Here, $b_h(M)$ is the Sheth-Tormen halo bias:
\be
b_h(M) = 1 + \frac{1}{\delta_c} \frac{d\log f}{d\log\sigma} \label{eq:sheth_tormen_halo_bias}
\ee
Note that $\sigma$, $n_h$, and $b_h$ are functions of both $M$ and $z$.

We assume that halos have NFW (Navarro-Frenk-White) density profiles~\cite{Navarro:1996gj}.
Recall that the NFW profile $\rho(r)$ has two parameters: the virial radius $r_{\rm vir}$ where the profile is
truncated, and the scale radius $r_s$ which appears in the functional form of the profile.
Sometimes, we reparameterize by replacing one of these parameters by the concentration $c = r_{\rm vir} / r_s$.
The NFW profile $u(r)$ and its Fourier transform $\tilde u(k)$ are given by:
\ba
u(r) &=& \frac{A}{(r/r_s) (1 + r/r_s)^2}  \hspace{1cm} (r \le r_{\rm vir})  \label{eq:nfw_real} \\
\tilde u(k) &=& 4\pi A r_s^3 \Bigg( - \frac{\sin(\kappa c)}{\kappa(1+c)} \nn \\
 && \hspace{0.5cm} + (\cos\kappa) \Big[ \Ci(\kappa(1+c)) - \Ci(\kappa) \Big] \nn \\
 && \hspace{0.5cm} + (\sin\kappa) \Big[ \Si(\kappa(1+c)) - \Si(\kappa) \Big] \Bigg)  \label{eq:nfw_fourier}
\ea
where $\kappa = k r_s$, and $\Si$ and $\Ci$ are the special functions:
\ba
\Si(x) &=& \int_0^x dt \, \frac{\sin t}{t} \\
\Ci(x) &=& -\int_x^\infty dt \, \frac{\cos t}{t} \nn \\
 &=& \gamma + \log(x) + \int_0^x dt \, \frac{\cos t - 1}{t}
\ea
and $\gamma = 0.577216\ldots$ is Euler's constant.
We choose the normalizing constant $A$ in Eqs.~(\ref{eq:nfw_real}),~(\ref{eq:nfw_fourier}) to be:
\be
A = \frac{1}{4\pi r_s^3} \left( \log(1+c) - \frac{c}{1+c} \right)^{-1}
\ee
With this value of $A$, the profile satisfies
$\tilde u(0) = \int_0^{r_{\rm vir}} dr \, (4\pi r^2) u(r) = 1$.

To use the NFW profile, we need expressions for the virial radius
$r_{\rm vir}(M,z)$ and halo concentration $c(M,z)$, as functions of
halo mass and redshift.
For the concentration, we use the fitting function from~\cite{Dutton:2014xda}:
\ba
\log_{10} c(M,z) &=& \alpha(z) + \beta(z) \log_{10}\left( \frac{M}{10^{12} \, h^{-1} M_\odot} \right)  \nn \\
\alpha(z) &=& 0.537 + 0.488 \exp\Big( -0.718 z^{1.08} \Big) \nn \\
\beta(z) &=& -0.097 + 0.024 z  \label{eq:dutton_maccio}
\ea
For the virial radius, we reparameterize by defining a virial {\em density}:
\be
\rho_{\rm vir} = \frac{3 M (1+z)^3}{4\pi r_{\rm vir}^3}  \label{eq:rho_r_vir}
\ee
then use the fitting function for $\rho_{\rm vir}$ from~\cite{Eke:1997ef}:
\ba
\rho_{\rm vir}(z) 
  &=& 178 \, \Omega_m(z)^{0.45} \rho_{\rm crit}(z) \nn \\
  &=& 178 \, \Omega_m(z)^{0.45} \left( \frac{3}{8\pi G} H(z)^2 \right)  \label{eq:eke_rvir}
\ea
The factor $(1+z)^3$ in Eq.~(\ref{eq:rho_r_vir}) arises because $\rho_{\rm vir}$
is a physical density, whereas $r_{\rm vir}$ is a comoving distance.

\subsection{Galaxies}
\label{app:gal_hod}

We assume that the number of galaxies in a halo of mass $M$ is
a Poisson random variable whose mean $\bN_g(M,z)$ is given by:
\be
\bN_g(M,z) = \left\{ \begin{array}{cl}
   (M/M_g(z)) & \mbox{if $M \ge M_g(z)$}  \\
      0  & \mbox{if $M < M_g$}
\end{array} \right.  \label{eq:gal_hod}
\ee
where $M_g(z)$ is the minimum halo mass needed to host a galaxy.

For each galaxy survey considered in this paper, we compute $M_g(z)$ by
matching to the redshift distribution $dn_g^{2d}/dz$, by numerically solving
the equation:
\be
\frac{dn_g^{2d}}{dz} = \Omega \frac{\chi(z)^2}{H(z)} \int_{M_g(z)}^{\infty} dM \, n_h(M) \frac{M}{M_g(z)}
\ee
for $M_g(z)$.  (This procedure for reverse-engineering a threshold halo mass $M_g(z)$
from an observed redshift distribution is sometimes called ``abundance matching''.)
The redshift distribution $dn_g^{2d}/dz$ is taken from~\cite{Sheldon:2011fm,Bilicki:2013sza,Aghamousa:2016zmz}
for SDSS-DR8, 2MPZ, and DESI-ELG respectively.
For each survey, the redshift distribution $dn_g^{2d}/dz$ and threshold halo mass $M_g(z)$
are shown in Figures~\ref{fig:dndz},~\ref{fig:mg}.

\subsection{FRB's}
\label{app:frb_hod}

Similarly, we model the FRB population by starting with a redshift
distribution $dn_f/dz$, which we take to be of the form:
\be
\frac{dn_f^{2d}}{dz} \propto z^2 e^{-\alpha z} \label{eq:dnf_dz}
\ee
for $0 \le z \le z_{\rm max}$, where the parameter $\alpha$ and maximum redshift $z_{\rm max}$ are given by:
\ba
\alpha &=& \left\{ \begin{array}{cl}
  3.5 & \mbox{(high-$z$ FRB model)} \\
  120 & \mbox{(low-$z$ FRB model)}
\end{array} \right.  \\
z_{\rm max} &=& \left\{ \begin{array}{cl}
  5  & \mbox{(high-$z$ FRB model)} \\
  0.12  & \mbox{(low-$z$ FRB model)}
\end{array} \right.
\ea
for our high-$z$ and low-$z$ fiducial FRB model respectively.
The FRB redshift distribution in both models is shown in Figure~\ref{fig:dndz}.

We assume that the number of FRB's in a halo of mass $M$ is a
Poisson random variable whose mean $\bN_f(M)$ is given by:
\be
\bN_f(M,z) = \left\{ \begin{array}{cl}
  \eta(z) \, (M/M_f)  & \mbox{if $M \ge M_f$}  \\
      0  &  \mbox{if $M < M_f$}
\end{array} \right.  \label{eq:frb_hod}
\ee
where $M_f$ is the threshold halo mass for hosting an FRB, and $\eta(z)$
is an FRB event rate per threshold halo mass.
In the FRB case, we take $M_f$ to be a free parameter, and determine $\eta(z)$
by abundance-matching to the FRB redshift distribution in Eq.~(\ref{eq:dnf_dz}).
In detail, we take:
\be
M_f = 10^9\ h^{-1} \ M_\odot
\ee
in both our high-$z$ and low-$z$ fiducial FRB models.
The prefactor $\eta(z)$ is then determined by numerically solving the equation:
\be
\eta(z) = \frac{dn_f^{2d}}{dz} \left( \Omega \frac{\chi(z)^2}{H(z)} \int_{M_f}^\infty dM \, n_h(M) \frac{M}{M_f(z)} \right)^{-1}
\ee
Thus, our FRB redshift distribution and HOD are parameterized by $(\alpha, z_{\rm max}, M_f)$, and the total
number of observed FRB's $N_f$ which determines the proportionality constant in Eq.~(\ref{eq:dnf_dz}).

We model dispersion measures by assuming that the host DM is
a lognormal random variable.  That is, the probability distribution is:
\be
p(D_h) = \frac{1}{D_h \sqrt{2\pi\sigma_{\log D}^2}} \exp\left( -\frac{(\log D_h - \mu_{\log D})^2}{2\sigma_{\log D}^2} \right)  \label{eq:dh_pdf}
\ee
where the parameters $(\mu_{\log D}, \sigma_{\log D})$ are given by:
\ba
\mu_{\log D} &=& \left\{ \begin{array}{cl}
  4 & \mbox{(high-$z$ FRB model)} \\
  6.78 & \mbox{(low-$z$ FRB model)}
\end{array} \right.  \\
\sigma_{\log D} &=& \left\{ \begin{array}{cl}
  1 & \mbox{(high-$z$ FRB model)} \\
  0.63 & \mbox{(low-$z$ FRB model)}
\end{array} \right.
\ea
The FRB DM distribution in both models is shown in Figure~\ref{fig:dndz}.

We assume that FRB's are observed with a Gaussian beam with FWHM $\theta_f$.
In the flat sky approximation, statistical errors on FRB location $(\theta_x, \theta_y)$
have the Gaussian probability distribution.
\be
p(\theta_x, \theta_y) = \frac{4 \log 2}{\pi \theta_f^2} \exp\left( -4 \log 2 \frac{\theta_x^2 + \theta_y^2}{\theta_f^2} \right)  \label{eq:frb_btheta}
\ee
By default, we take the FRB angular resolution to be $\theta_f = 1$~arcminute.

\subsection{Power spectra}

Given the model for halos, FRB's, and galaxies from the previous sections,
we are interested in angular power spectra of the form $C_l^{XY}$,
where each 2-d field $X,Y$ could be either a galaxy field (denoted $g$)
or an FRB field (denoted $f$).
We are primarily interested in cross power spectra $C_l^{fg}$, 
but auto spectra ($C_l^{ff}$, $C_l^{gg}$) also arise when forecasting 
signal-to-noise (e.g. Eq.~(\ref{eq:fisher_matrix})).

For maximum generality, we assume binned FRB and galaxy fields.
That is, the galaxy field is defined by specifying a redshift bin
$(z_{\rm min}, z_{\rm max})$, and keeping only galaxies which fall
in this range.
Similarly, the FRB field is defined by keeping only galaxies in
DM bin $(D_{\rm min}, D_{\rm max})$, after subtracting the galactic
contribution $\DM_{\rm gal}$.
Note that the unbinned galaxy field can be treated as a special case,
by taking the redshift bin large enough to contain all galaxies
(and analogously for the FRB field).

Before computing the power spectrum $C_l^{XY}$, we pause to
define some new notation.

For each tracer field $X$, let $\bN_X(M,z)$ denote the mean number
of tracers in a halo of mass $M$ at redshift $z$.
If $X$ is a binned galaxy field, in redshift bin $(z_{\rm min}, z_{\rm max})$, 
then $\bN_X(M,z)$ is given by:
\be
\bN_g(M,z) = \left\{ \begin{array}{cl}
    \frac{M}{M_g(z)} & \mbox{if $M \ge M_g(z)$ and $z \in [z_{\rm min}, z_{\rm max}]$} \\ [\medskipamount]
      0  & \mbox{otherwise.}
\end{array} \right.  \label{eq:galaxy_hod_binned}
\ee
generalizing Eq.~(\ref{eq:gal_hod}) for an unbinned galaxy field.
If $X$ is a binned FRB field, in DM bin $(D_{\rm min}, D_{\rm max})$, then:
\be
\bN_f(M,z) = \left\{ \begin{array}{cl}
\multicolumn{2}{c}{\eta(z) \frac{M}{M_f} \int_{D_{\rm min} - D_i(z)}^{D_{\rm max} - D_i(z)} dD_h \, p(D_h)}   \\  [\medskipamount]
   & \mbox{if $M \ge M_f$}  \\  [\bigskipamount]
      0 \phantom{XX} &  \mbox{if $M < M_f$}
\end{array} \right.  \label{eq:frb_hod_binned}
\ee
generalizing Eq.~(\ref{eq:frb_hod}) for an unbinned FRB field.
Here, $p(D_h)$ is the host DM probability distribution in Eq.~(\ref{eq:dh_pdf}),
and $D_i(z)$ is the IGM contribution to the DM at redshift $z$ (Eq.~(\ref{eq:dm_igm})).

For each tracer field $X$, let $n_X^{3d}(z)$ be the 3-d comoving number density,
and let $n_X^{2d}$ be the 2-d angular number density.  These densities can be written
explicitly as follows:
\ba
n_X^{3d}(z) &=& \int dM \, n_h(M) \bN_X(M,z) \\
n_X^{2d} &=& \int dz \, \frac{\chi(z)^2}{H(z)} n_X^{3d}(z)
\ea
Next, for a pair of tracer fields $(X,Y)$, let $n_{XY}^{2d}$ denote the angular
number density of object pairs $(x,y)$ which are co-located.
In our fiducial model, each FRB and galaxy is randomly placed within its halo,
so $n_{XY}^{2d}$ is zero unless the fields $X,Y$ contain the same objects.
That is, if the galaxy fields in non-overlapping redshift bins are denoted
$g_1, \cdots, g_M$, and the FRB fields in non-overlapping DM bins are denoted
$f_1, \cdots, f_N$, then:
\be
n^{2d}_{f_if_j} = n^{2d}_{f_i} \delta_{ij}
 \hspace{0.5cm}
n^{2d}_{g_ig_j} = n^{2d}_{g_i} \delta_{ij}
 \hspace{0.5cm}
n^{2d}_{f_ig_j} = 0  \label{eq:nfg_app}
\ee
One final definition.  For each tracer field $X$,
let $u^X_l(M,z)$ denote the angular tracer profile sourced by
a halo of mass $M$ at redshift $z$, normalized to $u=1$ at $l=0$.
The quantity $u^X_l(M,z)$ can be written explicitly as:
\ba
u^g_l(M,z) &=& \tilde u(M,k,z)_{k=l/\chi(z)}  \label{eq:ulg} \\
u^f_l(M,z) &=& b_l \tilde u(M,k,z)_{k=l/\chi(z)} \label{eq:ulf}
\ea
in the galaxy and FRB cases respectively.
Here, $\tilde u$ is the 3-d NFW profile in Eq.~(\ref{eq:nfw_fourier}), and
\be
b_l \equiv \exp\left( - \frac{\theta_f^2 l^2}{16 \log 2} \right)  \label{eq:bl_def}
\ee
is the Fourier-transformed FRB error distribution from Eq.~(\ref{eq:frb_btheta}).

Armed with the notation above, we can calculate the power spectrum $C_l^{XY}$
in a uniform way which applies to all choices of tracer fields $X,Y$.
The calculation follows a standard halo model approach, and we present
it in streamlined form.

Each tracer field $X$ is derived from catalog of objects at sky locations 
$\th^X_1, \cdots \th^X_N$.  The 2-d field $X$ is a sum of
delta functions in real space, or a sum of complex exponentials in Fourier space:
\ba
X(\th) &=& \frac{1}{n^{2d}_X} \sum_j \delta^2(\th-\th^X_j) \\
X(\l) &=& \frac{1}{n^{2d}_X} \sum_j e^{-i\l\cdot\th^X_j}
\ea
and likewise for $Y$.
The power spectrum $C_l^{XY}$ is defined by the equation:
\ba
\langle X(\l)^* \, Y(\l') \rangle
  &=& \frac{1}{n^{2d}_X n^{2d}_Y} \left\langle \sum_{jk} e^{i\l\cdot\th^X_j - i\l'\cdot\th^Y_k} \right\rangle  \nn \\
  &=& C_l^{XY} (2\pi)^2 \delta^2(\l-\l')  \label{eq:clxy_double_sum}
\ea
The double sum $\sum_{jk} (\cdots)$ can be split into three terms:
a sum over pairs $(j,k)$ of objects in different halos,
a sum over pairs $(j,k)$ of non-colocated objects in the same halo,
and a sum over co-located pairs $(j,k)$.
Correspondingly, the power spectrum $C_l^{XY}$ is the sum
of ``2-halo'', ``1-halo'', and ``Poisson'' terms:
\be
C_l^{XY} = C_l^{XY(2h)} + C_l^{XY(1h)} + C_l^{XY(p)}  \label{eq:clxy}
\ee
which are given explicitly as follows:
\ba
C_l^{XY(2h)} &=& \frac{1}{n_X^{2d} n_Y^{2d}} \int dz \frac{\chi(z)^2}{H(z)} n_X^{3d}(z) n_Y^{3d}(z) \nn \\
   && \hspace{0.2cm} \times b_X(z,l) b_Y(z,l) P_{\rm lin}(k,z)  \nn \\
C_l^{XY(1h)} &=& \frac{1}{n_X^{2d} n_Y^{2d}} \int dz \, dM \, \frac{\chi(z)^2}{H(z)} n_h(M,z) \nn \\
   && \hspace{0.2cm} \times \bN_X(M,z) \bN_Y(M,z) u_l^X(M,k) u_l^Y(M,k) \nn \\
C_l^{XY(p)} &=& \frac{n_{XY}^{2d}}{n_X^{2d} \, n_Y^{2d}}   \label{eq:clfg_master}
\ea
where in the first line we have defined:
\ba
b_X(z,l) &\equiv& \frac{1}{n_X^{3d}(z)} \int dM b_h(M,z) n_h(M,z) \nn \\
   && \hspace{0.2cm} \times \bN_X(M,z) u_l^X(M,z)  \label{eq:bias_X}
\ea
On large scales (where $u_l=1$), the quantity $b_X(z,l)$ reduces to the
bias parameter $b_X(z)$ defined in~\S\ref{sec:clfg}.

Throughout this paper, we have generally neglected the Poisson term
in $C_l^{fg}$, which arises if FRB's are actually located in survey
galaxies (in contrast to the 1-halo term, which arises if FRB's are in
the same halos as the survey galaxies).  This is equivalent to our
assumption in Eq.~(\ref{eq:nfg_app}) that $n^{2d}_{fg} = 0$.  If this
assumption is relaxed, then $C_l^{fg(p)}$ will be given by:
\be
C_l^{fg(p)} = b_l \frac{n_{fg}^{2d}}{n_f^{2d} \, n_g^{2d}}  \label{eq:clfg_poisson}
\ee
where the FRB beam convolution $b_l$ has been inserted by hand into the
general expression in Eq.~(\ref{eq:clfg_master}), since the FRB beam
displaces FRB's relative to their host galaxies.

\subsection{Free electrons}

When modeling propagation effects (\S\ref{sec:propagation_effects}), the 3-d
galaxy-electron power spectrum $P_{ge}(k,z)$ appears.  This can also be
computed in the halo model.

For simplicity, we will assume the approximation that all electrons are ionized.
This is a fairly accurate approximation: the actual ionization fraction is expected to
be $\approx 90\%$, with the remaining 10\% of electrons in stars, or ``self-shielding''
HI regions in galaxies.

We will also make the approximation that electrons have the same halo profiles as
dark matter.  This is a good approximation on large scales, but may overpredict $P_{ge}$
on small scales by a factor of a few.
This happens because dark matter is pressureless, whereas electrons have associated gas
pressure, which ``puffs out'' the profile.
In this paper our goal is modeling propagation effects at the order-of-magnitude level,
and it suffices to approximate electron profiles by dark matter profiles.
For a more precise treatment, fitting functions for electron profiles could be used \cite{Battaglia:2016xbi}.

Under these approximations, $P_{ge}$ is the sum $P_{ge} = P_{ge}^{1h} + P_{ge}^{2h}$
of one-halo and two-halo terms, given by:
\ba
P_{ge}^{1h}(k,z) &=& \frac{1}{\rho_{m,0} n_g^{3d}(z)} \int dM \, M n_h(M,z) \nn \\
 && \hspace{1.5cm} \times \bN_g(M,z) \tilde u(M,k,z)^2 \nn \\
P_{ge}^{2h}(k,z) &=& b_g(k,z) b_e(k,z) P_{\rm lin}(k,z)   \label{eq:pge}
\ea
where:
\ba
b_e(k,z) &=& \frac{1}{\rho_{m,0}} \int dM \, M b_h(M,z) n_h(M,z) \tilde u(M,k,z) \nn \\
b_g(k,z) &=& \frac{1}{n_g^{3d}(z)} \int dM \, b_h(M,z) n_h(M,z) \nn \\
  && \hspace{1.5cm} \times \bN_g(M,z) \tilde u(M,k,z)
\ea
Note that $b_e(k,z) \rightarrow 1$ as $k \rightarrow 0$.
Intuitively, the large-scale bias of free electrons is 1 in our model because
electrons perfectly trace dark matter ($\delta_e = \delta_m$).

\end{document}